\documentclass[usenatbib]{mnras}

\usepackage[T1]{fontenc}

\DeclareRobustCommand{\VAN}[3]{#2}
\let\VANthebibliography\thebibliography
\def\thebibliography{\DeclareRobustCommand{\VAN}[3]{##3}\VANthebibliography}

\usepackage{graphicx}
\usepackage{hyperref}
\usepackage{amsmath}
\usepackage{amssymb}
\usepackage{makecell}
\usepackage{comment}
\usepackage{xcolor}
\usepackage{enumerate}
\usepackage{threeparttable}
\usepackage[caption=false]{subfig}
\usepackage{calc}

\newcommand{\figurenv}[2]{\begin{figure}
\centering
\includegraphics[width=\linewidth]{plots/#1}
\caption{#2}
\end{figure}}

\newcommand{\pagefigurenv}[2]{\begin{figure*}
\centering
\includegraphics[width=\linewidth]{plots/#1}
\caption{#2}
\end{figure*}}

\newcommand{\doublefigenv}[7]{
\begin{figure*}
    \centering 
    \subfloat[#1]{%
      \includegraphics[width=#2\linewidth]{plots/#3}%
    }\hfill
    \subfloat[#5]{%
      \includegraphics[width=\linewidth-#2\linewidth]{plots/#6}%
    }
    \caption{
        (a) #4
        (b) #7
    }
    \end{figure*}
}

\def\rvir/{\ensuremath{R_{\mathrm{vir}}}}
\def\pgrav/{\ensuremath{P_{\mathrm{grav}}}}

\def\pturb/{\ensuremath{P_{\mathrm{disp}}}}
\def\pflow/{\ensuremath{P_{\mathrm{bulk}}}}
\def\ptherm/{\ensuremath{P_{\mathrm{therm}}}}
\def\pkin/{\ensuremath{P_\mathrm{kin}}}
\def\ptot/{\ensuremath{P_{\mathrm{tot}}}}

\def\pcold/{\ensuremath{P^\mathrm{cold}}} 
\def\pwarm/{\ensuremath{P^\mathrm{warm}}}
\def\phot/{\ensuremath{P^\mathrm{hot}}}

\def\pphasecold/{\ensuremath{\widetilde{P}^\mathrm{cold}_\mathrm{tot}}} 
\def\pphasewarm/{\ensuremath{\widetilde{P}^\mathrm{warm}_\mathrm{tot}}}
\def\pphasehot/{\ensuremath{\widetilde{P}^\mathrm{hot}_\mathrm{tot}}}

\def\rstarhalf/{\ensuremath{R_{*,1/2}}}
\def\ellpatch/{\ensuremath{\ell_\mathrm{patch}}}
\def\sigmasfr/{\ensuremath{\Sigma_\mathrm{SFR}}}

\def\mtwelvei/{\texttt{m12i}}
\def\mtwelvef/{\texttt{m12f}}
\def\mtwelveb/{\texttt{m12b}}

\def\th/{\ensuremath{^\mathrm{th}}}

\def\qeff/{\ensuremath{Q_\mathrm{eff}}}

\title[Pressure balance in the multiphase ISM of cosmologically simulated disk galaxies]{Pressure balance in the multiphase ISM of cosmologically simulated disk galaxies}

\author[Gurvich et al.]{
\parbox{\textwidth}{
Alexander B. Gurvich$^{1}$\thanks{E-mail: agurvich@u.northwestern.edu}, 
Claude-Andr\'e Faucher-Gigu\`ere$^{1}$, 
Alexander J. Richings$^{2}$,
Philip F. Hopkins$^{3}$, 
Michael Y. Grudi\'c$^{1}$, 
Zachary Hafen$^{1}$, 
Sarah Wellons$^{1}$, 
Jonathan Stern$^{1}$, 
Eliot Quataert$^{4}$, 
T.K. Chan$^{2,5}$, 
Matthew E. Orr$^{3}$, 
Du\v{s}an Kere\v{s}$^{5}$,
Andrew Wetzel$^{6}$, 
Christopher C. Hayward$^{7}$,
Sarah R. Loebman$^{6}$\thanks{Hubble Fellow},
and Norman Murray$^{8}$\thanks{Canada Research Chair in Theoretical Astrophysics}
}
\vspace{0.4cm}\\
\parbox{\textwidth}{
$^1${Department of Physics \& Astronomy and CIERA, Northwestern University, 1800 Sherman Ave, Evanston, IL 60201, USA}\\
$^2${Institute for Computational Cosmology, Durham University, South Road, Durham DH1 3LE, UK}\\
$^3${TAPIR, Mailcode 350-17, California Institute of Technology, Pasadena, CA 91125, USA}\\
$^4${Department of Astronomy and Theoretical Astrophysics Center, University of California Berkeley, Berkeley, CA 94720}\\
$^5${Department of Physics and Center for Astrophysics and Space Science, University of California at San Diego, 9500 Gilman Drive, La Jolla, CA 92093, USA}\\
$^6${Department of Physics, University of California, Davis, CA 95616, USA}\\
$^7${Center for Computational Astrophysics, Flatiron Institute, 162 Fifth Avenue, New York, NY 10010, USA}\\
$^8${Canadian Institute for Theoretical Astrophysics, 60 St. George Street, University of Toronto, ONM5S 3H8, Canada}\\
}}

\date{\vspace{-20pt}Submitted to MNRAS, May 2020}

\pubyear{2020}

\begin{document}
\label{firstpage}
\pagerange{\pageref{firstpage}--\pageref{lastpage}}
\maketitle

\begin{abstract}
    Pressure balance plays a central role in models of the interstellar medium (ISM), but whether and how pressure balance is realized in a realistic multiphase ISM is not yet well understood. We address this question by using a set of FIRE-2 cosmological zoom-in simulations of Milky Way-mass disk galaxies, in which a multiphase ISM is self-consistently shaped by gravity, cooling, and stellar feedback. We analyze how gravity determines the vertical pressure profile as well as how the total ISM pressure is partitioned between different phases and components (thermal, dispersion/turbulence, and bulk flows). We show that, on average and consistent with previous more idealized simulations, the total ISM pressure balances the weight of the overlying gas. Deviations from vertical pressure balance increase with increasing galactocentric radius and with decreasing averaging scale. The different phases are in rough total pressure equilibrium with one another, but with large deviations from thermal pressure equilibrium owing to kinetic support in the cold and warm phases, which dominate the total pressure near the midplane. Bulk flows (e.g., inflows and fountains) are important at a few disk scale heights, while thermal pressure from hot gas dominates at larger heights. Overall, the total midplane pressure is well-predicted by the weight of the disk gas, and we show that it also scales linearly with the star formation rate surface density ($\Sigma_{\rm SFR}$). These results support the notion that the Kennicutt-Schmidt relation arises because $\Sigma_{\rm SFR}$ and the gas surface density ($\Sigma_{\rm g}$) are connected via the ISM midplane pressure.
\end{abstract}



\section{Introduction}
\label{sec:introduction}
    
\subsection{ISM pressure balance and its connection to the regulation of star formation in galaxies}
Understanding the structure and dynamics of the interstellar medium (ISM) has long been recognized as a fundamental problem in astrophysics. 
Since the ISM is the reservoir out of which stars form, the phase structure, dynamics, and thermodynamics of interstellar gas directly affect the process of star formation \citep[e.g.,][]{2007ARA&A..45..565M, 2014PhR...539...49K}. 
Observations also show that basic ISM properties, including its metallicity, molecular gas content, mass in gas relative to stars, and velocity dispersion vary systematically as a function of galaxy parameters such as mass and redshift \citep[e.g.,][]{2011MNRAS.415...32S, 2020arXiv200306245T}. 
Thus, the physics of the ISM is also critical in understanding the formation and evolution of galaxies.

A principle central to essentially all models of the ISM, dating at least back to \cite{1956ApJ...124...20S}, is that of pressure balance between different phases.
This principle was a core component of the two-phase ISM model proposed by \cite{1969ApJ...155L.149F} and of the three-phase model of \cite{McKee1977}. 
In these classic analytic models, the pressure of each phase is assumed to be dominated by the thermal component, implying that $P_{\rm therm}=n k T$ is constant across phases. 
In the more modern picture, informed by a combination of observations and numerical simulations, the ISM is recognized as the host of many more complex processes. 
In particular, the structure and dynamics of the ISM are not only determined by thermal pressure but also by the mass, momentum, and energy stored and carried by turbulence \citep[e.g.,][]{2004ARA&A..42..211E, 2004RvMP...76..125M, 2012A&ARv..20...55H}. 
Moreover, observations show (and models predict) that in many regimes star-forming galaxies drive energetic galactic winds \citep[e.g.,][]{2017hsn..book.2431H, Fielding2018}. 
At their base, these galactic outflows can be viewed as flows which are not static but rather accelerated in the ISM. 
Thus, at the very least, a more complete understanding of the ISM should involve \emph{dynamic} equilibrium \citep[][]{Kim2011, Sun2020}.

\pagefigurenv{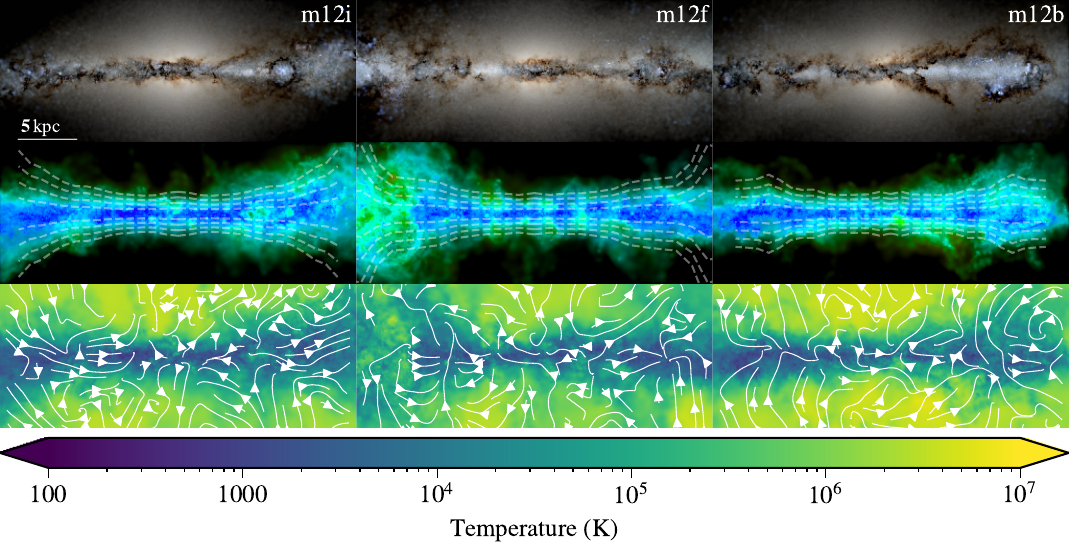}{
	\label{f:pretty_m12i} Renderings of Milky Way-mass galaxies from the FIRE-2 cosmological zoom-in simulations, \mtwelvei/, \mtwelvef/, and \mtwelveb/ at redshift $z=0$ viewed edge-on. 
    \textit{Top row:} A rendering of the mock stellar light, where pixel colors are determined by brightness in the \textit{ugr} bands and including dust extinction, where emission in each band is determined by stellar age and metallicity and dust abundance is proportional to the gas metallicity. 
    \textit{Middle row:} A projection of the gas density and temperature wherein the saturation of each pixel is proportional to the log of the column density and the hue is proportional to the log of the temperature.
    The result is an image where bright and blue pixels contain dense and cold gas while fuzzy and green pixels contain gas that is diffuse and hot. 
    Dashed contours are plotted at integer values of the measured gas scale height ($\pm h$, $2h$, $3h$, and $4h$; defined in \S \ref{s:disk_characteristics}). 
    \textit{Bottom row:} Gas projection through a thin slice ($-250 \leq y \leq 250$ pc) colored by temperature to indicate  the cold ($T\leq 10^3$ K), warm ($10^3 \leq T \leq 10^5$ K), or hot ($T\geq 10^5$ K) phase.
    Gas velocity streamlines are plotted in white to give an impression of the complex vertical velocity structure.}
    
    Beyond the structure of the ISM, our interest in pressure balance is motivated by the role this principle may play in regulating star formation on galactic scales. 
    Galaxies are observed to exhibit a tight correlation between their globally averaged gas surface density, $\Sigma_\mathrm{g}$, and star formation rate surface density, $\Sigma_{\mathrm{SFR}}$  \citep[e.g.,][]{Kennicutt1998,Genzel2010,Kennicutt2012}.
    This correlation is referred to as the Kennicutt-Schmidt (KS) relation. 
    Several analytic models have been developed in which the KS relation arises as a result of a balance between feedback and gravity \citep[e.g.,][]{Thompson2005,Ostriker2010,Ostriker2011,Faucher-Giguere2013,Hayward2017}. 
    Although the detailed assumptions vary between different implementations, the common basis for this class of ``equilibrium'' models is that the ISM pressure at the midplane must balance the weight of the overlying gas, $P_{\rm mid}\sim \Sigma_{\rm disk} \Sigma_\mathrm{g}$ (where $\Sigma_{\rm disk}=\Sigma_\mathrm{s} + \Sigma_\mathrm{g}$ is the total mass surface density). 
    Assuming that $P_{\rm mid} \propto \Sigma_{\rm SFR}$, which can be motivated in scenarios in which the energy necessary to maintain the ISM pressure is provided either by stellar feedback or by the release of gravitational energy by inspiraling gas \citep[e.g.,][]{Krumholz2018}, this directly predicts a relation of the form $\Sigma_{\rm SFR} \propto \Sigma_{\rm disk} \Sigma_\mathrm{g}$.
    
    While the equilibrium KS models are appealing for their simplicity and ability to explain basic observations, it is not \textit{a priori} guaranteed that the model assumptions (e.g. that a steady state is reached, that vertical pressure balance is achieved, or that pressure is correlated with the star formation rate) are realized in actual galaxies. 
    As mentioned above, real galaxies have a dynamic, multiphase ISM shaped by a combination of gravitational, cooling, and stellar feedback processes. 
    Additionally, in a realistic cosmological environment, galaxies continuously accrete new gas from the intergalactic medium (IGM) and frequently experience interactions (including mergers) with nearby galaxies. 
    Moreover, many galaxies drive powerful outflows that eject gas from the ISM into the IGM \citep{Steidel2010,Muratov2015}. 
    These complexities are typically neglected in existing analytic models \citep[see][for an exception]{Hayward2017}, but their effects can be investigated using simulations which include some or all of these processes. 
    The focus of this paper is on dissecting the structure of and analyze vertical pressure balance in the ISM of simulations which are fully cosmological yet of sufficiently high resolution to capture its multiphase nature. 
    Also presented is a brief analysis of the validity of equilibrium models for explaining the KS relation that emerge in such simulations; this will be expanded and built upon in future papers.
    
    \pagefigurenv{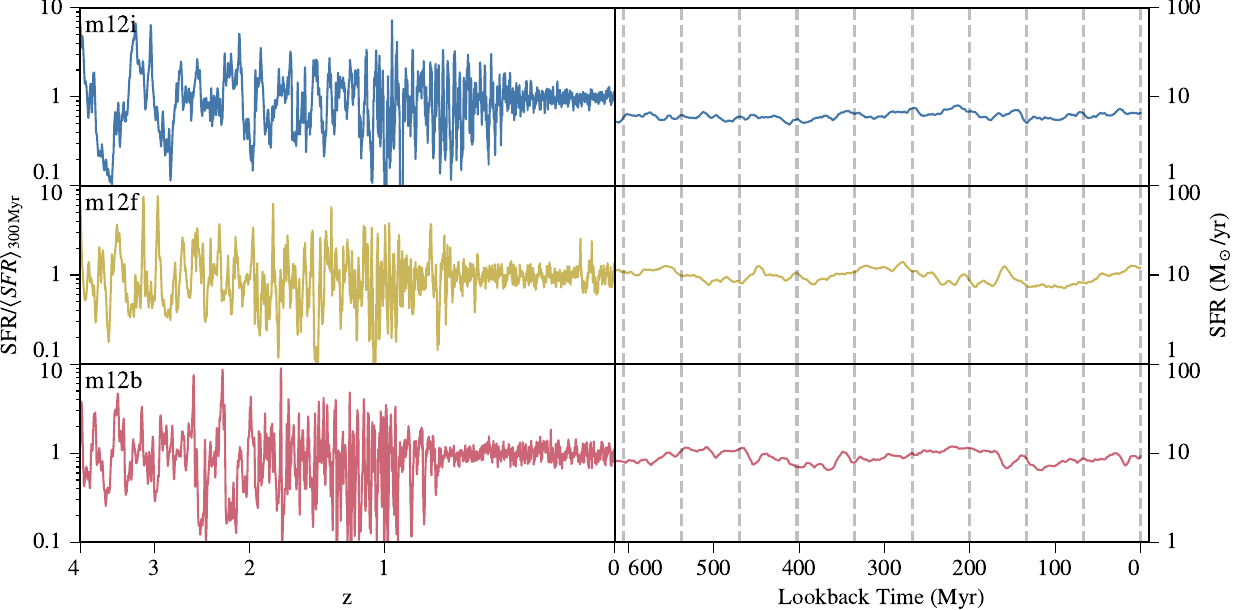}{
        \label{f:galaxies_sfh} \emph{Left column:} Star formation histories for our three simulated galaxies, normalized by the 300 Myr running average.
        Although the SFR is highly variable at high redshift, these galaxies all undergo a significant transition at $z\leq 1$, following which the SFR becomes much less bursty and a stable, well-ordered, galactic disk forms.
        \emph{Right column:} Star formation histories for the same galaxies over the past 600 Myr before redshift $z=0$, which is the time window on which the analysis in this paper focuses.  
        Vertical dashed lines are plotted to show the times corresponding to the main snapshots that we use for this analysis. 
        The star formation rates are time steady (modulo minor stochastic fluctuations) 
        over the course of the 600 Myr window that we investigate.}

\subsection{Previous numerical work}
    There has been significant previous work testing the assumption of vertical pressure balance in numerical simulations, which has been critical in establishing the present evidence in support of equilibrium models of the KS relation. 
    However, previous analyses have relied on various assumptions or simplifications. While simplified setups have the advantage of enabling some parameters of the problem to be better controlled, it is also important to relax some of the assumptions in order to develop a more complete understanding of the physics. 
    For example, a large number of previous studies were based on high-resolution local simulations of ${\sim}1$ kpc patches of galactic disks with idealized boundary conditions \cite[e.g.,][]{Kim2011,2012ApJ...754....2S,Kim2013,Kim2015,Vijayan2020}. 
    In addition to neglecting interactions with the cosmological environment, it has been shown that local Cartesian box simulations cannot correctly model the development of galactic winds at large heights from the disk \citep{Martizzi2016,Fielding2017}. It is unclear whether this has implications for pressure balance at the galaxy-wind interface. 
    Some studies of ISM pressure balance have also been carried out using global simulations of isolated galaxies \citep[][]{Benincasa2016,Benincasa2019TheGalaxies}. 
    These analyses of global galaxy models broadly supported the results of local-box simulations that the ISM tends to settle into a statistical steady state whose vertical structure is determined by a balance between pressure and gravity. 
    However, these simulations of isolated galaxies neglected the cosmological context. 
    Moreover, both local-box and isolated galaxy simulations previously analyzed for vertical pressure balance implemented simplified models of stellar feedback. 
    For instance, the simulations referenced above modeled supernovae (SNe) and/or photoelectric heating by far-ultraviolet (FUV) radiation, but did not explicitly include stellar winds, radiation pressure on dust, or model the disruption of molecular clouds by HII regions. 
    Since the assumptions for stellar feedback can have large effects on the dynamics and phase structure of the ISM, it is important to explore different, and also more complete, sets of feedback models.
    
\subsection{This paper}
    In this work, we investigate ISM pressure balance and its implications for the KS relation using the FIRE-2 simulations from the FIRE project \citep[][]{Hopkins2014, Hopkins2018}\footnote{FIRE project website: \url{http://fire.northwestern.edu}}. 
    These are high-resolution, fully cosmological, zoom-in simulations which implement a detailed model for the ISM, star formation, and stellar feedback.
    The result is a dynamic, multiphase ISM self-consistently shaped by gravity, cooling, and feedback.
    The FIRE-2 simulations have been shown to provide a good match to several basic observed properties of galaxy populations, including the stellar mass-halo mass relation \citep[][]{Hopkins2014, Hopkins2018}, the mass-metallicity relation \citep[][]{2016MNRAS.456.2140M}, galactic winds \citep[e.g.,][]{Muratov2015,2017MNRAS.470.4698A}, lifetimes and properties of giant molecular clouds \citep{2019arXiv191105251B,2020MNRAS.492..488G}, and the KS relation \citep{Orr2018}.

    We focus the present analysis on a set of low-redshift, Milky Way-mass simulated galaxies from the ``Latte'' suite of simulations \citep{Wetzel2016}. 
    In FIRE, galaxies in this regime have large, long-lived galactic disks and relatively steady star formation rates.
    We systematically analyze the breakdown of the total ISM pressure into different gas phases (cold/warm/hot) and different forms of pressure (thermal/dispersion/bulk flow, which we define in \S \ref{s:hydro_balance_equation}). 
    By comparing the pressures in different phases/forms to the weight of overlying gas we quantify the extent to which vertical pressure balance is achieved. 
    We defer an analysis of galaxies in which star formation is highly time variable, or ``bursty,'' which is predicted in lower mass galaxies and at high redshift in the FIRE simulations \citep[e.g.,][]{Sparre2017, Faucher-Giguere2018}, to future work. 
    We also note that this paper focuses on a theoretical analysis of the (thermo) dynamical properties of the ISM in the simulated galaxies. 
    In particular, we do not attempt detailed comparisons with observations, which will be explored in future work. 
    This is both because the simulation results are of interest in and of themselves, and also because proper comparisons with ISM and star formation observations require detailed modeling of gas and star formation tracers. 
    For a much more detailed assessment of how the FIRE simulations compare with observations of the KS relation, we refer to \cite{Orr2018}. 

    The outline of this paper is as follows.
    In \S \ref{s:disk_characteristics} we describe the simulations analyzed in this paper and the basic properties of their galactic disks, including their Toomre $Q$ parameter. 
    In \S \ref{s:hydro_balance_equation} we outline our methods for measuring pressure contributions and vertical pressure balance in our simulations.  
    We present our main results in \S \ref{s:hydro_balance_results}, where we quantify ISM pressure balance as a function of averaging scale, galactocentric radius, height from the midplane, gas phase, and form of pressure.
    We discuss our results, including in the context of equilibrium models for the origin of the KS relation, in \S \ref{s:discussion}. 
    We summarize our results in \S \ref{s:conclusion}. 
    Appendices summarize some numerical tests and methods.
    
\vspace{-0.15in}
\section{Simulations and Basic Disk Properties}
\label{s:disk_characteristics}
    \subsection{Simulations}
    \label{sec:simulations}
    The simulations used in our analysis are FIRE-2 cosmological zoom-in simulations and are run in the Meshless Finite Mass (MFM) mode of the GIZMO\footnote{Information about GIZMO and a public version of the code is available at: \url{https://www.tapir.caltech.edu/~phopkins/Site/GIZMO}} gravity+magnetohydrodynamic code. 
    MFM is a Lagrangian, mesh-free, finite-mass method which combines advantages of traditional smooth particle hydrodynamics (SPH) and grid-based methods
    \citep[for numerical details and tests, see][]{Hopkins2015}. 
    The FIRE-2 physics model, described in full detail in \cite{Hopkins2018}, 
    includes radiative cooling for gas down to $10$ K (including an approximate treatment of fine-structure metal and molecular lines).  
    Star particles that represent simple stellar populations are formed in gas that is self-gravitating, dense ($n_\mathrm{H}\geq 1000$ cm$^{-3}$), and molecular \citep[the importance of these different criteria is discussed in][]{2013MNRAS.432.2647H}. 
    When all criteria for star formation are satisfied, the dense gas is converted into stars with 100\% efficiency per local free-fall time. 
    Thus, a low star formation efficiency on galactic scales is not ``put in by hand'' but rather emerges as a result of regulation by stellar feedback \citep[][]{Hopkins2014, Orr2018}.\footnote{It is possible that processes other than stellar feedback, e.g. turbulence sourced by gravity, also play a role in the regulation of galaxy-scale star formation. However, feedback is essential since simulated galaxies based on the same physics (including gravity, cooling, turbulence, etc.) but neglecting stellar feedback experience runaway ISM collapse and SFRs that exceed observations by 1-2 orders of magnitude \citep[e.g.,][]{Hopkins2011}.} 
    
    Star particles return mass, metals, momentum, and energy into the ISM, using rates that are functions of the star particle's age following the \texttt{STARBURST99} population synthesis model \citep{Leitherer1999}. 
    These feedback processes include supernovae (Type II and Ia), stellar winds from O, B, and AGB stars, photoelectric heating, photoionization, and radiation pressure. 
    The model for radiation pressure includes both short-range and long-range components \citep[][]{2020MNRAS.491.3702H}; the long-range component in particular can contribute spatially smooth pressure supporting the overlying ISM which is not accounted for in the kinematic motions of the gas. 
    We will return to this point when interpreting pressure balance near the midplane (\S \ref{s:hydro_balance_results}). 
    
    In this paper we focus on three representative Milky Way-mass galaxies from the FIRE project with halo masses ${\sim}10^{12}$ M$_\odot$ near redshift $z=0$: \mtwelvei/, \mtwelvef/, and \mtwelveb/ \citep[see][for detailed discussion of each of these simulations]{Wetzel2016,Garrison-Kimmel2017,Garrison-Kimmel2019}.
    Though the GIZMO code includes the capability to model magnetic fields and cosmic rays, the simulations we analyze in this paper do not include these physics (we briefly comment on the role of cosmic rays in \S \ref{s:pism_mid}).
    In FIRE, these simulations form large, long-lived central disks at $z\lesssim 1$ \citep[][]{2018MNRAS.481.4133G}. 
    The main galaxy in \mtwelvef/~experiences a significant merger at $z\approx 0.1$, but this occurs $\approx 600$ Myr before the beginning of the time window on which we focus our analysis (see \S \ref{s:disk_properties} below).  
    The resolution of the baryons in these simulations is $m_\mathrm{b}\approx 7100$ M$_\odot$ for gas cells and star particles. 
    Thus, GMCs with masses $M_{\rm GMC}\sim 10^{6}-10^{7}$ M$_{\odot}$ contain $\gtrsim 100-1000$ resolution elements.
    Gravitational softening lengths for gas cells are adaptive and reach values of ${\sim}5$ pc at the average density for star formation \citep[see Table 3 of][]{Hopkins2018}, so massive star-forming complexes are also well resolved spatially.
    \begin{table*}
    \caption{\label{t:sim_table} Properties of the simulated galaxies analyzed in this paper}
    \centering
    \begin{threeparttable}
        \centering
        \begin{tabular}{c | c | c| c | c | c}
            Name & $M_\mathrm{halo}$ ($10^{12}$ M$_\odot$) \tnote{a} & \rstarhalf/ (kpc) \tnote{b}& $M_*$ ($10^{10}$ M$_\odot$) \tnote{c} & $f_\mathrm{g}$ \tnote{d} & $\langle SFR \rangle_\mathrm{600\,Myr}$ (M$_\odot$ yr $^{-1}$) \tnote{e}\\ \hline
            m12i & 1.1 & 2.9 & 6.7 & 0.20 & 6.2\\
            m12f & 1.5 & 4.0 & 8.5 & 0.24 & 9.7\\
            m12b & 1.3 & 2.8 & 8.8 & 0.16 & 9.0\\
        \end{tabular}
        \begin{tablenotes}
            \item[a] The total dark matter and baryonic mass within \rvir/, the virial radius defined as in \cite{Bryan1998}.
            \item[b] The 3D radius containing half of the total stellar mass within 15\% of $R_\mathrm{vir}$.
            \item[c] The total stellar mass within the galaxy (defined as within 5\rstarhalf/ cylindrically and within $\pm 5$ kpc of the midplane).
            \item[d] $f_\mathrm{g} = \frac{M_\mathrm{g}}{M_*+M_\mathrm{g}}$ with $M_\mathrm{g}$ the total gas mass within the galaxy. 
            \item[e] Average star formation rate in the 600 Myr period before redshift $z=0$.
        \end{tablenotes}
        \end{threeparttable}
    \end{table*}
    
    Figure \ref{f:pretty_m12i} shows $z=0$ edge-on renderings of mock stellar emission (including dust extinction), density-temperature projections, and temperature slice-projections for each of the three galaxies.
    All three simulations show relatively thin (${\sim}\,400 \mathrm{\,pc} - 1$ kpc) disks with multiphase gas distributions.
    Figure \ref{f:galaxies_sfh} shows the star formation history of these three galaxies and illustrates that the star formation rates are time-steady at $z\sim0$, despite much more bursty earlier histories.
    
    We note that the average SFRs in the present simulations are somewhat higher than the Milky Way \citep[${\sim}1.7$ M$_{\odot}$ yr$^{-1}$;][]{2016ARA&A..54..529B}
    because (1) they have higher gas masses and (2) the Milky Way itself has an SFR below the typical $L_*$ galaxy today by a factor of 2-3 \citep[e.g.][]{2015ApJ...809...96L}. 
    Therefore, while the simulated galaxies are hosted by Milky Way-mass dark matter halos, they are not perfect Milky Way analogues. 
    This does not affect the validity of our results, since we are interested in the general physical processes that regulate star formation in disk galaxies and not in reproducing the detailed properties of our Galaxy.

    \pagefigurenv{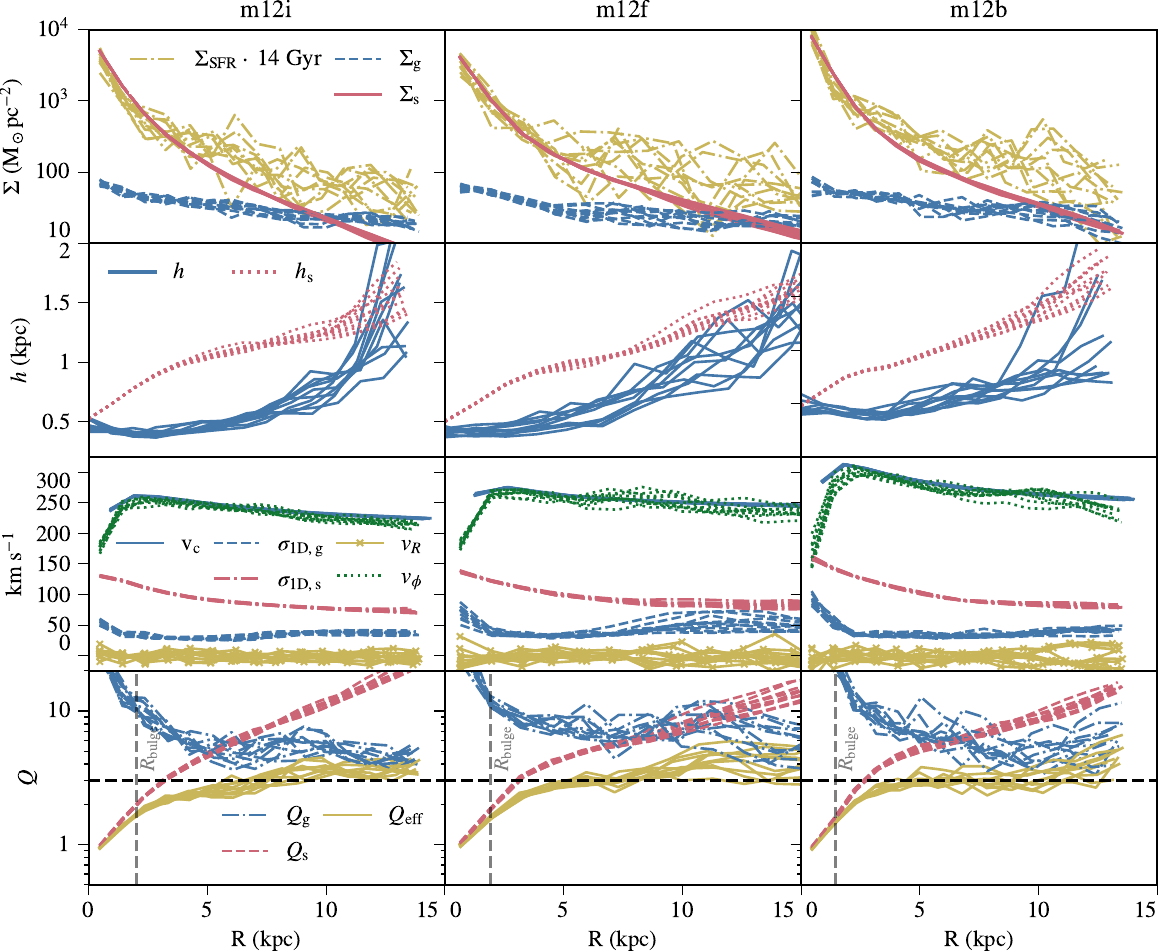}{
        \label{f:radial_profiles} 
        Disk properties vs. cylindrical radius $R$ in annuli of width $\rstarhalf/ / 3\sim 1$ kpc in \mtwelvei/, \mtwelvef/, and \mtwelveb/ over 10 snapshots separated by $\approx 60$ Myr at redshift $z\sim 0$.
        The bulge radius, estimated as the mass-weighted 3D radius of the counter-rotating stars, is plotted as a vertical dashed line in the bottom row.
        \textit{First row:} Surface densities of gas mass ($\Sigma_\mathrm{g}$), stellar mass ($\Sigma_s$), and star formation rate (\sigmasfr/).
        \textit{Second row:} Gas and stellar scale heights ($h$ and $h_s$ respectively).
        \textit{Third row:} Circular velocity, gas and stellar velocity dispersions ($\sigma_{1D,g}$ and $\sigma_{1D,s}$ respectively), gas radial velocity ($v_R$), and gas azimuthal velocity ($v_\phi$).
        \textit{Fourth row:} Effective Toomre $Q$ parameter ($\qeff/$) combining Toomre parameters evaluated for gas ($Q_\mathrm{g}$) and stars ($Q_\mathrm{s}$) separately.
        Each line is a profile from a single snapshot and the relatively small variation between lines indicates the galaxies are nearly in steady state. 
        }
        
    \subsection{Disk properties}
    \label{s:disk_properties}
    We locate the main galaxy in each zoom-in simulation output using the Amiga Halo Finder \citep[AHF;][]{Gill2004,Knollmann2009}. 
    We define halos using the overdensity criterion \citep[following][]{Bryan1998}, which corresponds to $\approx 100\times$ the critical density at $z=0$. 
    We define the location of the ``main galaxy'' to be the halo center of mass (computed using all of the dark matter, stellar, and gas components). 
    Since we limit our analysis to times during which the disks are stable and do not experience major mergers, we find the halo center of mass to be a sufficiently accurate definition of galaxy center. 
    In our analysis, we focus on galactocentric radii $\leq 5\rstarhalf/$, where \rstarhalf/ is the radius that contains half of the total stellar mass that lies within 15 \% of the virial radius (\rvir/). 
    This outer radius corresponds to the data that we process our analysis pipeline; for many plots we focus on smaller radii for which the results are most relevant.  
     Table \ref{t:sim_table} summarizes these, and other, important properties for the three simulations we use.
    
    We define the vertical ($z$) axis to lie along the total angular momentum vector of the gas contained within $5\rstarhalf/$.
    Since, in principle, the stellar and gas disks can be misaligned we have checked that our results are insensitive to the choice between these options. 
    We have also confirmed that there is no significant warping of the disk in the simulation snapshots that we analyze by comparing the angle between the angular momentum vectors of radial annuli (finding differences $<1^{\circ}$). 
    The $x-y$ (or $R-\phi$) plane is normal to the $z$-axis, and the midplane, corresponding to $z=0$, is defined to intersect the halo center of mass. 
    With this orientation, we measure the gas disk scale height $h$ as a function of cylindrical galactocentric radius $R$, for different annuli of width $\rstarhalf//3$, by fitting an exponential profile of the form $\rho(z) \propto e^{-|z|/h}$ to the vertical distribution of gas within $\pm$ 3 kpc of the midplane. 
    We fit independent profiles both above and below the midplane and, though these heights are typically very similar, use the average of the two scale heights to define a single $h$ that we use for the rest of the analysis. 
    
    Figure \ref{f:radial_profiles} catalogs important radius-dependent properties of the three galaxies in our sample including: the gas and stellar surface densities ($\Sigma_\mathrm{g}$, $\Sigma_\mathrm{s}$ respectively), the star formation rate surface density (\sigmasfr/), the gas and stellar scale heights ($h$, $h_\mathrm{s}$), the circular velocity ($v_c=\sqrt{G M_{\rm tot}(<r)/r}$), the one-dimensional velocity dispersions of the gas and stars ($\sigma_\mathrm{1D,g}$, $\sigma_\mathrm{1D,s}$), the mass-weighted mean gas radial and azimuthal velocities ($v_R$, $v_\phi$), and Toomre $Q$ parameters (see \citeauthor{Orr2018} \citeyear{Orr2018} for detailed comparisons of $\Sigma_g$ and $\sigmasfr/$ in FIRE simulations to observations of the KS relation). 
    
    For all surface densities, quantities are integrated vertically between $\pm 5h$, and for all velocities, quantities are measured using resolution elements between $\pm h$.
    Different curves represent the profiles from different snapshots in time, separated by $\approx 60$ Myr and spanning $\approx 600$ Myr, and overlap strongly (except at the outermost radii), reflecting the stability of the disks during the analysis period. 
    We repeated the analysis presented here and in the following sections with $3\times$ as many snapshots separated by $\approx 20$ Myr spanning the same $600$ Myr period and find that all of our time-averaged results are well converged with the $60$ Myr time-spacing (see Appendix \ref{a:time_variability}).
    
    The velocity dispersions $\sigma_\mathrm{1D}$ for both the gas and stars are defined as the average width of the velocity distributions in the $r,\phi,$ and $z$ directions measured as if the distribution were Gaussian and the 84\th/ and 16\th/ percentiles corresponded to the $\pm\sigma$ range i.e.
    \begin{equation}
        \sigma_\mathrm{1D}^2 = \frac{1}{3}\sum_{i=r,\phi,z}\left(\frac{p_{84}(v_{i}) - p_{16}(v_{i})}{2}\right)^2.
    \end{equation}
    We find that the gas 1D-velocity dispersion is roughly constant with radius at ${\sim} 30$ km s$^{-1}$, excluding the inner-most region (which is likely affected by  the presence of a stellar bulge), while the stellar 1D-velocity dispersion decreases substantially with radius.
    The gas 1D-velocity dispersion is only modestly anisotropic, with $2/3 \sigma_{\mathrm{g},r} \approx \sigma_{\mathrm{g},\phi} \approx \sigma_{\mathrm{g},z}$ uniformly with radius.
    The annulus-averaged radial velocity is small, ${\sim} 2-3$ km s$^{-1}$ in magnitude with both positive and negative direction, indicating there is not a strong net axisymmetric inflow of gas.
    The azimuthal velocity is nearly the circular velocity for all except the innermost radii \citep[see][ for a discussion of deviations of $v_\phi$ from $v_c$, and their implications, in the FIRE simulations]{Wellons2020}.
    
    \subsection{Toomre Q}
    We also measure the the Toomre $Q$ parameter in annuli, a proxy for the stability of a rotating disk against gravitational collapse  \citep[e.g.][]{Toomre1964,Elmegreen2011}.
    To characterize our disks, which include both gas and stars, we follow \cite{Wang1994} who define an ``effective'' $Q$ that includes the effects of a two-component disk (gas+stars).
    \begin{equation}
    \label{e:Qeff}
        {\qeff/}^{-1} = {Q_\mathrm{s}}^{-1} + {Q_\mathrm{g}}^{-1},
     \end{equation}
    where $Q_\mathrm{g}$ and $Q_\mathrm{s}$ are defined for the gas and stars separately:
     \begin{equation}
            Q_\mathrm{g}=\frac{\Omega \sigma_\mathrm{1D,g}}{\pi G \Sigma_\mathrm{g}} \qquad \mathrm{and}\qquad
            Q_\mathrm{s}=\frac{\Omega \sigma_\mathrm{1D,s}}{\pi G \Sigma_\mathrm{s}},
    \end{equation}
    and $\Omega$ is the orbital frequency of the system as a whole. 
    
    The bottom panel in Figure \ref{f:radial_profiles} shows the effective Toomre $Q$ parameters as a function of radius in our three simulations. 
    We find that $\qeff/$ is ${\sim} 3$ for radii that are outside the influence of the galactic bulge  and increases to ${\sim} 10$ for radii greater than $15$ kpc (not shown), where there is little star formation, for all three simulations. 
    We return to and interpret this result in \S \ref{s:discussion}. 

\section{Equations of Vertical Pressure Balance}
\label{s:hydro_balance_equation}
    \figurenv{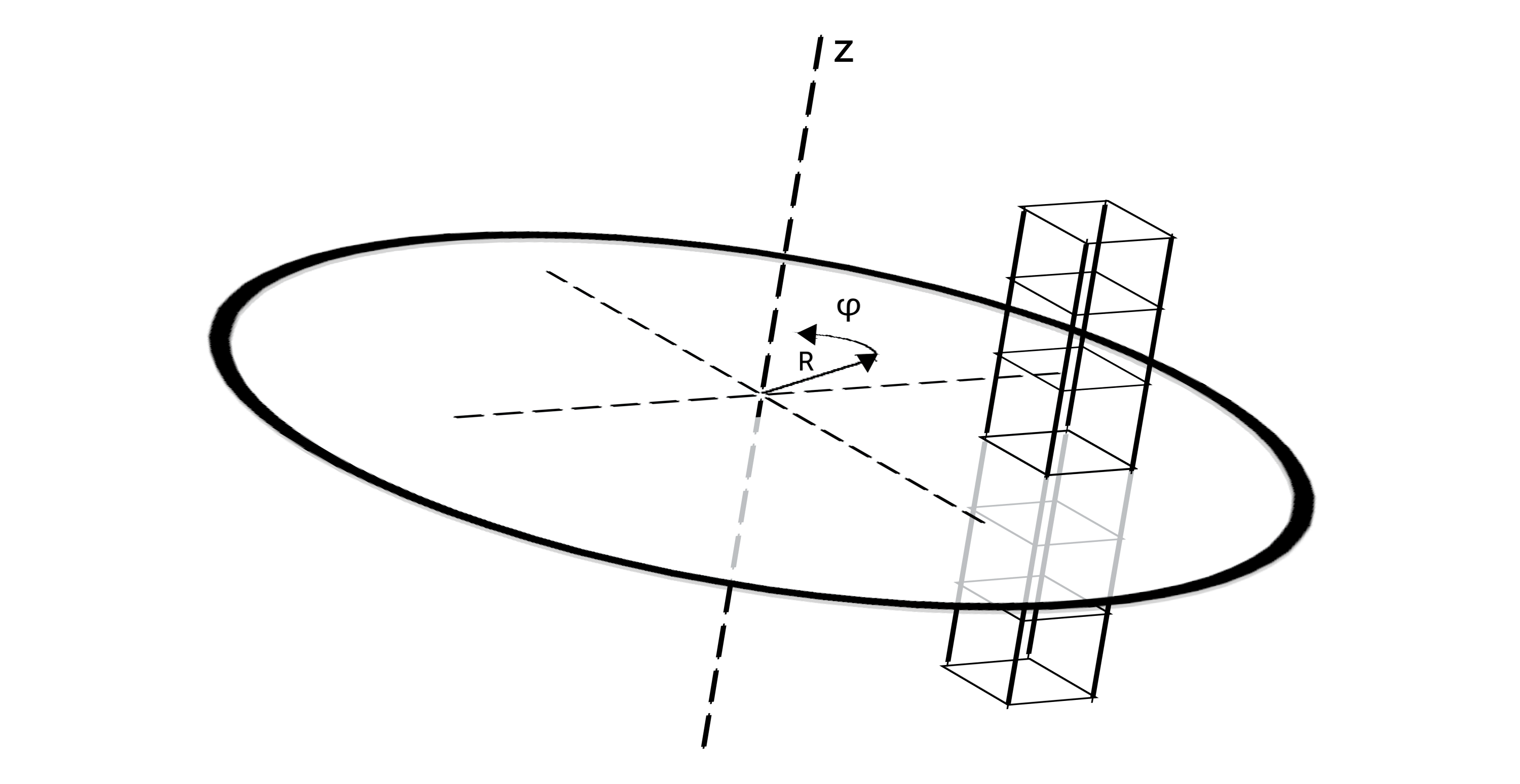}{
        \label{f:disk_schematic} A schematic of the analysis setup to illustrate the orientation of our coordinate system.
        The $z-$direction is defined to lie along the axis of the gas angular momentum vector, the cylindrical $R$ direction increases outwards from the barycenter of the dark matter halo, and the $\phi$ direction increases in the direction of galactic rotation. 
        An example column of ``slabs,'' in which we measure quantities, are drawn at an arbitrary position in the disk.}
    
    \subsection{Definition of `columns' and `slabs'}
    Here we define a framework within which we measure the physical properties of different ISM phases, with the goal of testing whether (and how) vertical pressure balance is realized in different regions of the galaxy.
    To do this, we subdivide the galaxy within $5$\rstarhalf/ into square $x-y$ ``patches'' (or apertures) of size $\ell_{\mathrm{patch}}\times \ell_{\mathrm{patch}}$, where $\ellpatch/=$100 pc, 300 pc, 1 kpc, and 3 kpc is varied to study the dependence of the results on averaging scale. 
    These values for \ellpatch/ are representative of some spatially resolved galaxy observations \citep[e.g.][]{Leroy2008, Gallagher2018,Sun2020}; we do not expect any of our results to depend significantly on the exact shape of the patches used to define the analysis. 
    Each patch is assigned a scale height based on the galactocentric radius of the patch center (as described in \S \ref{s:disk_properties} above). 
    Each patch is then subdivided vertically into ``columns'' oriented parallel to the $z-$axis of 400 ``slabs'' of equal thickness from $-20h \to 20h$.
    Thus, in each column there are 200 slabs above the midplane and 200 slabs below the midplane, each of width $h/10$. 
    The overall geometry of patches, columns, and slabs thus defined is illustrated in Figure \ref{f:disk_schematic}.
    
    \vspace{-0.15in}
    \subsection{Generalized `dynamic' pressure balance}
    \label{sec:generalizedz_balance}
        In cosmological simulations, the ISM is not necessarily in \textit{hydrostatic} balance at all times and locations because of (i) density and velocity fluctuations; and (ii) bulk flows such as fountain flows above (or below) the disk plane that can contribute significantly to supporting the overlying gas. 
    There can also be external perturbations, due to e.g. satellite galaxies. 
    We therefore derive more general relationships between averaged kinematic/dynamic/thermodynamic galactic properties implied by the general momentum (Euler) fluid equation:
    
    \begin{equation}
        \label{e:general_Euler}
        \left(\frac{\partial}{\partial t} + \vec{u}\cdot \vec\nabla\right)\vec u  = -\frac{1}{\rho}\vec\nabla P + \vec g + \vec S,
    \end{equation}
    where $\vec u$ is the velocity field, $P$ is the total effective pressure, and $\vec g$ is the gravitational acceleration (due to the total mass in the system).  
    The $\vec S$ term on the right-hand side corresponds to momentum injection (expressed in units of acceleration), such as by stellar feedback. 
    In many circumstances, we may expect that the average momentum injection rate is locally zero, e.g. because the vector sum of momentum injected by a point-like supernova explosion is zero by construction in our subgrid model \citep[][]{2018MNRAS.477.1578H}. However, some momentum injection is explicitly non-local (including the long-range radiation pressure in the FIRE model) and even nominally ``local'' feedback injection can locally average to a non-zero net momentum because, numerically, the momentum is injected over a finite volume consisting of ${\sim}32$ neighboring MFM cells. 
    We thus keep track of the momentum source term $\vec S$ in the derivation below.
    
    Next, we note that our primary interest is pressure support on scales $z\sim h$, where $|z| \ll r$ for thin disks. 
    On these scales, we assume that the system in each slab is well approximated by a steady state with plane-parallel geometry, i.e. we only consider the $z-$component of Equation (\ref{e:general_Euler}). 
    With these approximations, and after multiplying both sides by $\rho$, the $z$-equation can be written in cylindrical coordinates as
    \begin{align}
        \label{e:Euler_z}
        \rho \left(\frac{\partial u_z}{\partial t} +
        u_R\frac{\partial u_z}{\partial R} + 
        \frac{u_\phi}{R}\frac{\partial u_z}{\partial \phi} + 
        u_z\frac{\partial u_z}{\partial z}\right)& \nonumber \\ 
        = -\frac{\partial P}{\partial z}+ \rho g_z& + \rho S_z.
    \end{align}
    Let us now focus on the $u_{z} \partial u_{z} / \partial z$ term on the left-hand side. 
    Using the mass continuity equation,
    $\partial \rho / \partial t +  \vec \nabla \cdot ( \rho \vec u ) = 0$, 
    it is straightforward to verify that
    
    \begin{align} 
        &\rho u_z\frac{\partial u_z}{\partial z} = \notag \\
        &u_z \frac{\partial \rho}{\partial t} + 
        u_z \frac{\partial \left( \rho u_R\right)}{\partial R} +
        \frac{\rho u_R u_z}{R} +
        \frac{u_z}{R} \frac{\partial \left( \rho u_\phi\right)}{\partial \phi} + 
        \frac{\partial \left(\rho u_z^2\right)}{\partial z}
    \end{align}
    
    \noindent so that, after combining terms with like partial derivatives, Equation (\ref{e:Euler_z}) can be expressed as
    \begin{align} 
        \label{e:Euler_z_mass_cont}
        &\frac{\partial \left(\rho u_z\right)}{\partial t} + 
        \frac{\partial \left( \rho u_R u_z \right)}{\partial R} + 
        \frac{\rho u_R u_z}{R}+ \nonumber \\
        &\frac{1}{R} \frac{\partial \left( \rho u_\phi u_z\right)}{\partial \phi} + 
        \frac{\partial \left(\rho u_z^2 + P\right)}{\partial z}
        = \rho g_z + \rho S_z
    \end{align}
    
    Since we focus our analysis on time periods during which the simulated disks are long-lived, with approximately constant properties, we assume that the disks are well described by a \textit{statistical} steady state. 
    At any given time, the detailed instantaneous properties of the galaxy (e.g., the turbulent gas density and velocity fields) can differ significantly from the statistical average.
    Under this assumption, the $\partial/\partial t = 0$ on the left-hand side of Equation (\ref{e:Euler_z_mass_cont}) can be neglected after time averaging:\footnote{Note that upon time averaging between an initial time $t_{i}$ and a final time $t_{f}$, $\langle \partial(\rho u_z)/\partial t \rangle \propto \int_{t_i}^{t_f} dt' \partial(\rho u_z)/\partial t' = \Delta (\rho u_z) |^{t_{f}}_{t_{i}}$. 
        Even in a statistical steady state, the instantaneous value of $\rho u_z$ can vary significantly from $t_{i}$ to $t_{f}$ within a single column.  
        However, after averaging over many independent columns, the fluctuations should cancel. 
        We show in Appendix \ref{a:time_variability} that our results are indeed insensitive to this term.}
    
    \begin{align}
        \label{e:Euler_z_mass_cont_ss}
        &\left \langle \frac{\partial \left( \rho u_R u_z \right)}{\partial R} \right \rangle + 
        \left \langle \frac{\rho u_R u_z}{R} \right \rangle \nonumber\\ 
        &\left \langle \frac{1}{R} \frac{\partial \left( \rho u_\phi u_z\right)}{\partial \phi} \right \rangle +
        \left \langle \frac{\partial \left(\rho u_z^2 + P\right)}{\partial z} \right \rangle  \nonumber \\
        &= \left \langle \rho g_z \right \rangle + \left \langle \rho S_z \right \rangle &
    \end{align}
    
    Later in this paper, we plot results up to $|z|=10h$ to show how the flow develops as the disk joins the CGM. 
    At these large $|z|$, the plane-parallel geometry assumption can break down significantly and so we do not necessarily expect Equation (\ref{e:Euler_z_mass_cont_ss}) to be accurately satisfied. 
    
    We then \emph{define} the rest frame of each column such that 
    it comoves with an annulus at radius $R$. 
    Specifically, we give each column an azimuthal velocity equal to the (mass-weighted) average disk gas azimuthal velocity at $R$ and a radial velocity equal to the average radial velocity at the same radius in annuli of width $\rstarhalf/ / 3\sim 1$ kpc
    (the same annuli used in \S \ref{s:disk_characteristics} and Figure \ref{f:radial_profiles}). 
    Because of this choice, the column rest frames are generally not inertial and the Euler equation should include terms corresponding to fictitious centrifugal and Coriolis forces. 
    However, in plane-parallel geometry these fictitious terms act purely in the $x-y$ plane and so they do not modify Equation (\ref{e:Euler_z}). In the rest frame of any given column, we therefore have $\langle u_R \rangle=0$ and $\langle u_\phi \rangle=0$ by construction. 
    
    Our final assumption is that the $R$ and $\phi$ components of the velocity field are statistically independent of the $z$ component and the density field $\rho$, so that the time averages of terms involving products of $R$ and $z$ velocity components, or $\phi$ and $z$ velocity components, factor out\footnote{This is not strictly guaranteed but is a reasonable simplifying assumption given the complex (in some respects turbulent) dynamics of the ISM.}:

    \begin{align}
    \label{e:factor_vels}
    \left \langle \frac{\partial \left( \rho u_R u_z \right)}{\partial R} \right \rangle  &= 
        \frac{\partial }{\partial R} \langle u_R \rangle \langle \rho u_z \rangle = 0, \\ \notag 
    \left \langle \frac{\rho u_R u_z}{R} \right \rangle  &= \frac{1}{R} \langle  u_R \rangle \langle \rho u_z \rangle = 0, \\ \notag 
    \left \langle \frac{1}{R} \frac{\partial \left( \rho u_\phi u_z\right)}{\partial \phi} \right \rangle &= 
        \frac{1}{R}\frac{\partial }{\partial \phi }\langle u_\phi \rangle  \langle \rho u_z \rangle = 0.
    \end{align}
    This leaves us with the following simplified equation:  
    \begin{equation}
        \label{e:Euler_z_mass_cont_ss_boost}
        \left\langle \frac{\partial (P + \rho u_{z}^{2})}{\partial z}
       \right\rangle
         = \left \langle \rho g_z \right \rangle + \left \langle \rho S_z \right \rangle
    \end{equation}
    
    \figurenv{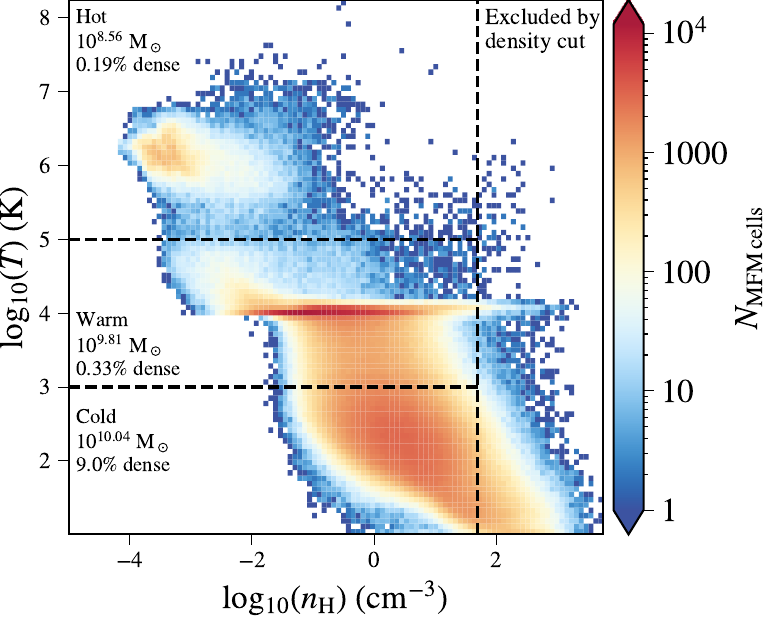}{
        \label{f:temperature_density} A temperature-density diagram of the particles within $5\rstarhalf/$ cylindrically and within $\pm 20$ kpc from the midplane vertically in \mtwelvei/ at redshift $z=0$.
        Horizontal lines are plotted at $T=10^3$ K and $T=10^5$ K to denote the warm-cold and warm-hot boundaries respectively. 
        Likewise, a vertical line is plotted at a density of $50$ cm$^{-3}$ to denote the cut above which we excise dense gas so as to exclude clumps that do not participate in vertical pressure balance.
        The percentage of MFM cells in each phase that are classified as ``dense'' (and thus excluded), along with the total mass in each phase, is annotated on the plot. }
        
    We now take a coarse-grained view by averaging both sides of Equation (\ref{e:Euler_z_mass_cont_ss_boost}) over the volume of a slab.
    The superscript `$s$' distinguishes slab-averaged quantities from their continuously defined values. 
    There is an important subtlety in the definitions that follow. 
    Although the volume average of Equation (\ref{e:Euler_z_mass_cont_ss_boost}) is evaluated without any weighting, we define several of the slab-averaged quantities as weighted by mass, e.g. $u_{z}^{s}= \langle u_{z} \rangle_{\rm M}$, $g_{z}^{s}= \langle g_{z} \rangle_{\rm M}$, $S_{z}^{s}= \langle S_{z} \rangle_{\rm M}$, ..., where $\langle x \rangle_{\rm M} = (\int dV \rho x)/m_{\rm slab}$. 
    The slab density is defined as $\rho^{s} = m_{\rm slab}/V_{\rm slab}$. 
    Since the different terms in Equation (\ref{e:Euler_z_mass_cont_ss_boost}) involve a factor of density $\rho$, this yields simple (but fully rigorous) expected relationships between slab-averaged quantities. 
    In this coarse-grained view, $P^{s}$ is the sum of the usual thermal term and a ``dispersion'' term (which is often referred to as the ``turbulent pressure,'' but which can contain motions which are not strictly turbulent) corresponding to the kinetic energy of flows that are resolved by the simulation but on scales $<\ellpatch/$: $P^{s} = \ptherm/^{\rm s} + \pturb/^{\rm s}$, where $\ptherm/^{\rm s} = (3/5) \rho^{s} c_{s}^{2}$, $\pturb/^{\rm s} = \rho^{s} \sigma_{z}^{2}$, $c_{s}^{2}$ is the mass-averaged sound speed squared,\footnote{Note that $\ptherm/^{\rm s} = \rho^{\rm s} \langle (v_{\rm th,z} - u_{z}^{s})^{2}\rangle_{\rm M}=(3/5) \rho^{\rm s} c_{\rm s}^2$ for a monatomic gas with adiabatic index $\gamma=5/3$, where $v_{\rm th,z}$ is the $z$ component of thermal velocities.}  and $\sigma_{z}^{2} = \langle (u_{z} - u_{z}^{s})^{2} \rangle_{\rm M}$.
    Finally, we define a ``bulk flow'' pressure $\pflow/^{\rm s} \equiv \rho^{s} {u_{z}^{s}}^{2}$, which accounts for momentum transfer owing to the net flow through a slab. 

    After these definitions, we have:\footnote{Since eq. (\ref{e:Euler_z_mass_cont_ss_boost}) involves a partial derivative with respect to $z$ on the left-hand side, which interacts with the volume average, the slab-averaged version should involve $z$ ``boundary terms,'' similar to the $\Delta \rho u_{z}$ that we neglected when taking the time average of eq. (\ref{e:Euler_z_mass_cont}). In this case, the boundary terms cancel exactly in the limiting of infinitely thin slabs, and we show in Appendix \ref{a:time_variability} that our results are also insensitive to these boundary terms.} 
    \begin{align}
    \label{e:final_balance}
       \left\langle \frac{\partial \ptot/^s}{\partial z}
       \right\rangle  = \left\langle \rho^{s} g_z^{s} \right\rangle + \left \langle \rho^{s} S_z^{\rm s} \right \rangle,
    \end{align} 
    where 
    \begin{align}
       \ptot/^{\rm s} = \ptherm/^{\rm s} + \pturb/^{\rm s} + \pflow/^{\rm s}.
    \end{align}
    Note that, since $g_z^{s}$ and $S_z^{s}$ are defined as mass-weighted averages over slabs, it is the case that $[\rho g_z]^{\rm s} = \rho^{\rm s} g_z^{\rm s}$ and $[\rho S_z]^{\rm s} = \rho^{\rm s} S_z^{\rm s}$, a fact used on the right-hand side of Equation (\ref{e:final_balance}).
    
    We can then integrate each side from height $z$ above the midplane to  $\infty$. 
    We assume that $\ptot/(\infty)=0$, which is a good assumption so long as the midplane pressure is much larger than the pressure out in the CGM. 
    Integrating the left-hand side gives the slab-averaged pressure $P_{\rm tot}^{\rm s}(z)$, while the right-hand side gives the weight per unit area of the gas above the slab at height $z$, if $z>0$:
    \begin{equation}
        \label{e:hb_above}
        W_\mathrm{tot}^{\rm s}(z) \equiv - \int^{\infty}_z dz' \
        \rho^{s}(z') g_z^{s}(z').
    \end{equation}
    \noindent with an analogous integral with bounds $-\infty\to z$ if $z<0$ (below the midplane).
    Neglecting the momentum source term, as is commonly done in similar analyses \citep[e.g.,][]{Kim2015}, the expression for steady state balance in Equation (\ref{e:final_balance}) is then equivalent to $\langle P_{\rm tot}^{\rm s}(z) \rangle = \langle W_{\rm tot}^{\rm s}(z) \rangle$ for all $z$. 
    In addition to testing this result against the simulations, in this work we analyze how $P_{\rm tot}$ is partitioned into thermal, dispersion, and bulk flow, as a function of $R$ and $z$. 
    
    \pagefigurenv{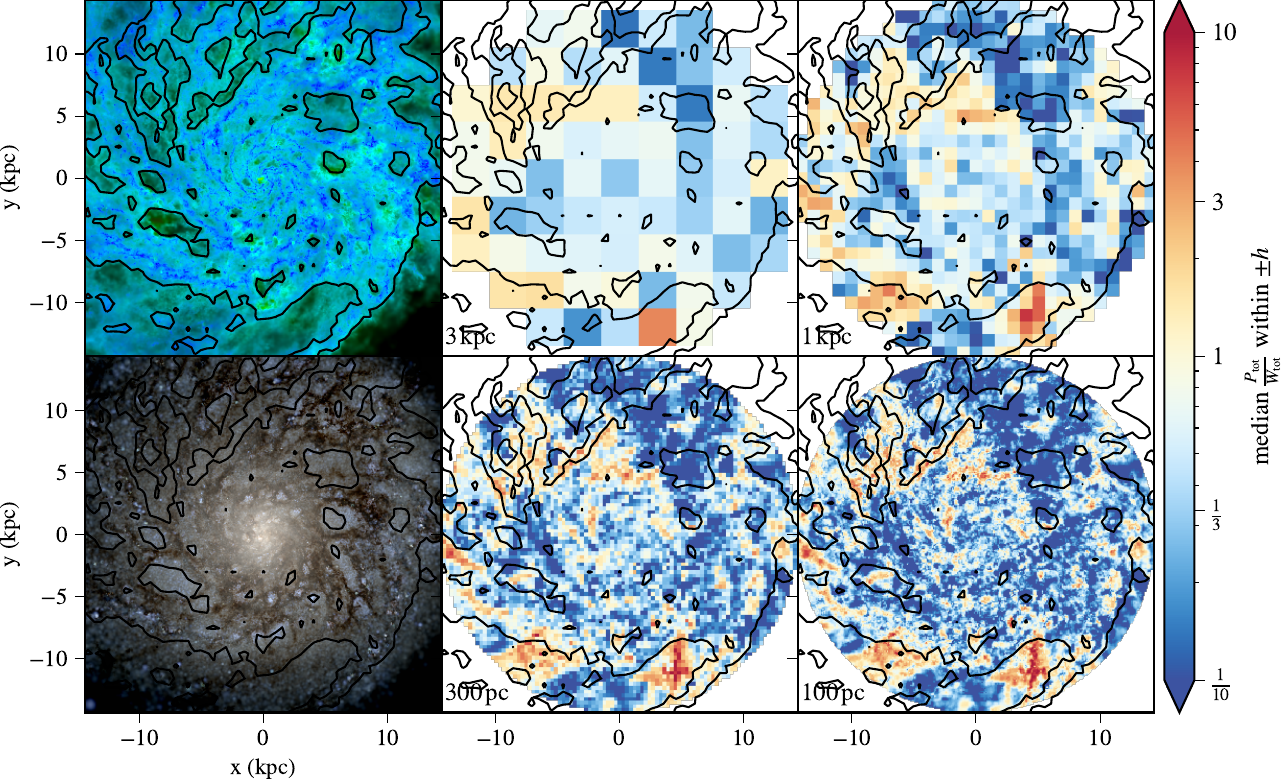}{
        \label{f:pressure_ratio_map_scales} 
        The two columns on the right-hand side show the median pressure-to-weight ratio within $|z|\leq h$, including all phases, at redshift $z=0$ in \mtwelvei/. 
        Different panels show different patch sizes \ellpatch/, denoted in the bottom left-hand corner of each panel. 
        Face-on renderings of the gas (top left-hand panel) and stars (bottom left-hand panel) analogous to edge-on renderings in Figure \ref{f:pretty_m12i} are provided as a reference for spiral and other high-density features. 
        The contours correspond to gas surface density $\geq 10$ M$_\odot$/pc$^2$ and are intended to aid comparison between panels.
        In general, deviations from vertical pressure balance increase with decreasing patch size (from the top middle panel to the bottom right-hand panel) owing to increasing scatter. 
        The majority of the volume is occupied by under-dense and under-pressurized gas whereas over-pressurized gas correlates with heating by stellar feedback around actively star-forming regions. 
        }
        
    In practice, to compute slab-averaged quantities, we smoothly deposit gas properties onto the grid of slabs in an SPH-like way (see Appendix \ref{a:sph_smooth}). 
    To evaluate $W_{\rm tot}$, we use \texttt{pykdgrav}\footnote{
        \texttt{pykdgrav} is a PYTHON package for efficiently computing $N-$body gravitational accelerations (and potentials) using a k-d tree. See \url{https://github.com/mikegrudic/pykdgrav/}.} 
    to compute the gravitational acceleration in the $z$-direction, taking into account all the particles within the virial radius of the main halo. 
    To minimize noise, 
    we average four evaluations of $g_z(z)$, each at a point $\ellpatch//4$ away from the patch center and along a different cardinal direction. 
    This procedure effectively replaces $g_{z}^{\rm s}$ (defined as a mass-weighted average) by a slab-volume averaged version; we expect this to be a minor effect as the gravitational acceleration is dominated by long-range forces except in the densest gas, which is excluded from the analysis as described in the next section.
    
\pagefigurenv{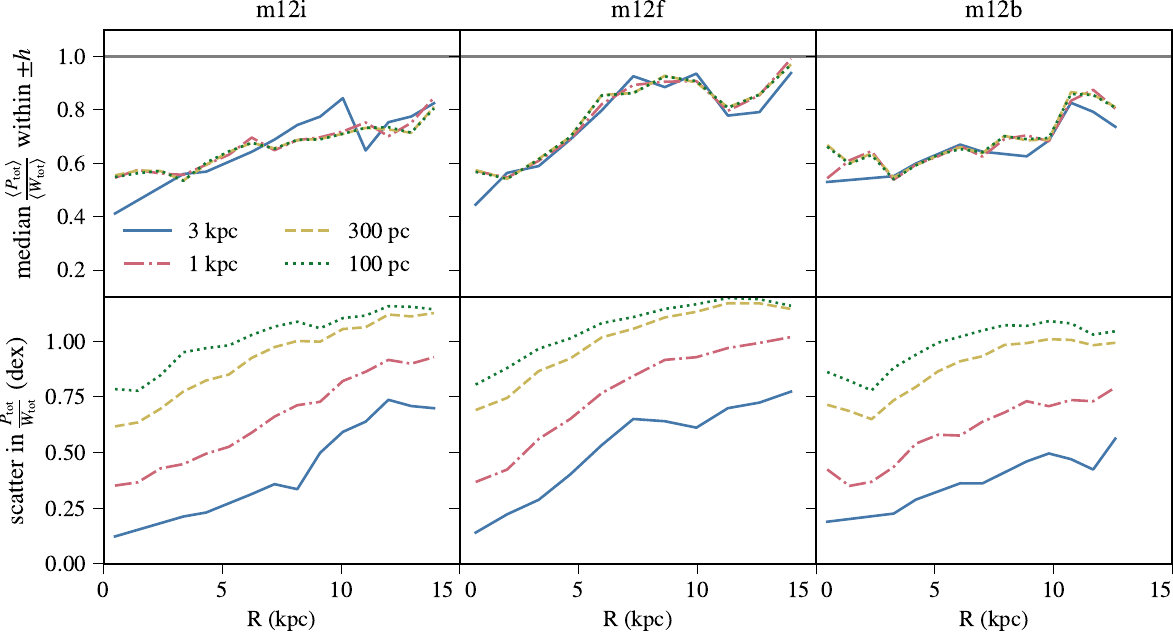}{
        \label{f:HB_scatter_vs_radius} 
        Radial profiles of the ``height-median pressure ratio'' (top row) and the ``height-median patch-to-patch scatter'' (bottom row) of the pressure ratio at redshift $z\sim 0$ and averaged over $\Delta t\approx 600$ Myr (approximately the ratio of the midplane pressure to the vertical weight of gas in the disk, see \S \ref{sec:PW_vs_r_z} for exact definitions).
        Both are evaluated using slabs within $\pm h$ of the midplane.
        Different patch sizes, \ellpatch/, are plotted in different line styles ranging from 3 kpc to 100 pc. 
        Each column corresponds to a different simulation, indicated by the top headings.
        Overall, the height-median pressure systematically increases from ${\sim}0.5$ to ${\sim}1$ with increasing galactocentric radius, independent of the averaging scale.
        On the other hand, the patch-to-patch scatter increases both with radius and with decreasing averaging scale.
   }
   
\subsection{Partial vs. local pressures in different phases}
    \label{s:partial_vs_local_P}
    To understand how pressure support is partitioned in a multiphase ISM, we also divide the gas in the galaxy into three phases according to its temperature: cold ($T<10^3$ K), warm ($10^3<T<10^5$ K), and hot ($T>10^5\,\mathrm{K}$).
    In addition to this temperature cut, we exclude all gas denser than $50$ cm$^{-3}$ to excise very dense clumps that are largely decoupled from the volume-filling flow and effectively do not participate in vertical pressure balance.
    We find that our results are insensitive to density cuts above $10$ cm$^{-3}$, below which substantial amounts of diffuse cold and warm gas are erroneously excised. 
    Figure \ref{f:temperature_density} shows a temperature-density diagram illustrating the different gas phases in one of our simulations.
    
    In simplified models, such as the classic three-phase model of the ISM \citep[e.g.][]{McKee1977}, the different phases are assumed to have the same ``local'' pressure, i.e.
    \begin{equation}
        \pphasecold/ \approx \pphasewarm/ \approx \pphasehot/,
    \end{equation}
    whereas here we use the term ``local'' to refer to the pressure ($\propto$ energy density) in the volume \textit{occupied by a given phase}.
    We contrast this with the ``partial'' pressure, which we define as the volume weighted contribution to the total volume integrated pressure.
    The partial and local pressures for each phase are then related by the volume-filling fraction of a phase, $f_\mathrm{phase} = V_\mathrm{phase}/V_\mathrm{slab}$. 
    
    \begin{equation}
        \label{e:local_pressure}
         P^\mathrm{phase} = f_\mathrm{phase}\widetilde{P}^\mathrm{phase}.
    \end{equation}
    The total ISM pressure can then be written as the sum of the total partial pressures from each phase.
    \begin{equation}
        \label{e:total_pressure_phases}
        \ptot/ = \pcold/_\mathrm{tot} + \pwarm/_\mathrm{tot} + \phot/_\mathrm{tot}
    \end{equation}
    
    In the rest of this paper, we use the ``tilde'' notation as above to distinguish between local vs. partial pressures.
  
 \pagefigurenv{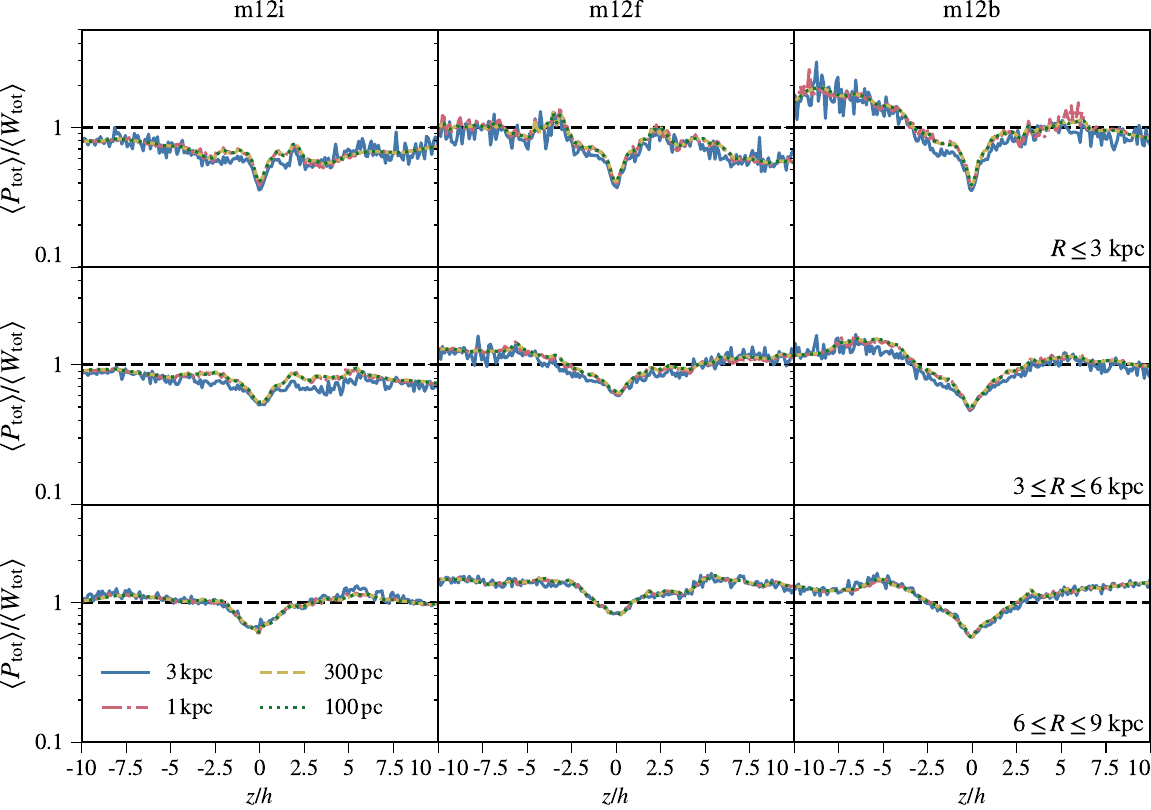}{
    	\label{f:HB_vertical_profiles_sims} 
        Vertical profiles of the annulus-averaged pressure-to-weight ratio at redshift $z\sim 0$ and averaged over the last $\Delta t\approx 600$ Myr as a function of vertical position $z$ in units of the scale height $h$ (see Figure \ref{f:radial_profiles} for scale height as a function of radius). 
        Different patch sizes, \ellpatch/, are plotted in different line styles ranging from 3 kpc to 100 pc. 
        Rows divide patches into radial annuli (denoted in the bottom right-hand corner of the panels). In general, the pressure-to-weight ratio is $\approx 1$, representing approximate vertical pressure balance, with only small variation between different \ellpatch/. 
        Systematic departures near the midplane are evident in all three simulations analyzed, indicating the importance of additional terms in equation (\ref{e:final_balance}) (see \S \ref{sec:PW_vs_r_z} for further discussion). 
        }
        
    \doublefigenv{partitioning pressure between phases \label{f:pressure_ratio}}{0.4175}{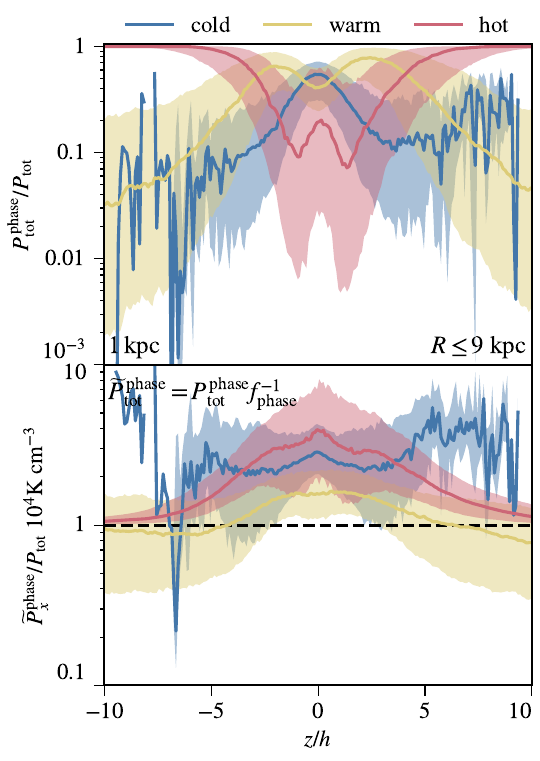}{
        \label{f:pressure_partition}
        \emph{Top panel:} Partial pressures for the cold ($T\leq 10^3$ K), warm ($10^3 \leq T \leq 10^5$ K), and hot ($T\geq 10^5$ K) gas phases, normalized by the total ISM pressure \ptot/  as a function of height from the disk midplane.  
        For each phase, all pressure components (thermal, dispersion, and bulk flow) are included. 
        \emph{Bottom panel}: ``Local'' pressure in each phase ($\widetilde{P}$, defined in Equation (\ref{e:local_pressure}), also including all components) relative to \ptot/. 
        These pressures differ from the partial pressures in the top panel by the inverse of the volume filling fraction $f_{\rm phase}$. 
        Ratios ${\sim} 1$ implies pressure equilibrium between phases.
        }
        {partitioning pressure between components \label{f:pressure_term_fractions}}{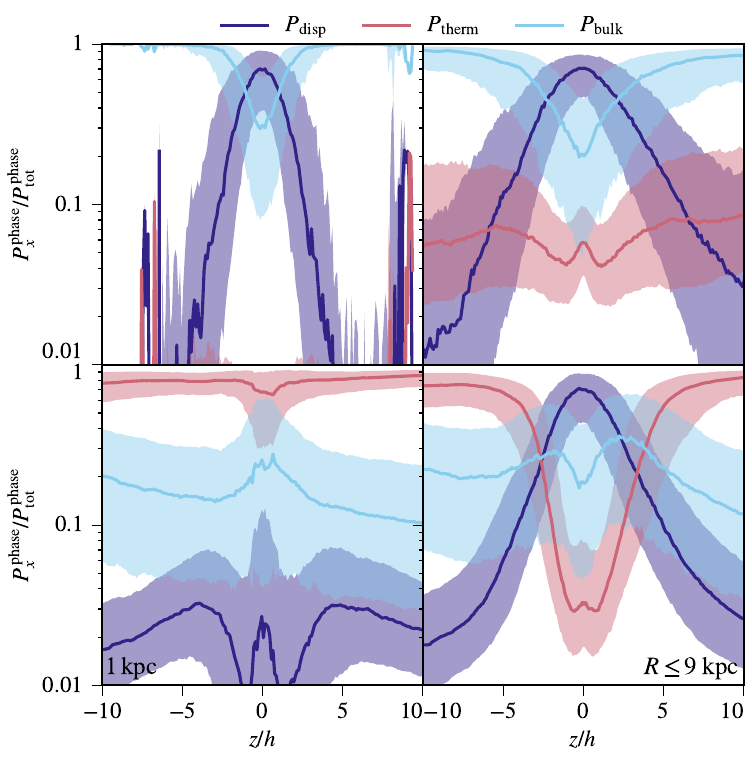}{
        Fractional contributions for each pressure component (thermal, dispersion, or bulk flow) to the total pressure in each gas phase (top left: cold, top right: warm, bottom left: hot, bottom right: all). 
        For the cold phase, the thermal contribution is so small that it is below the $y-$axis. 
        In each panel, solid curves show patch-medians and shaded regions are the inter-quartile ranges, which quantify patch-to-patch scatter.
        These results are evaluated for \mtwelvei/ in patches with $R\leq 9$ kpc, using a fiducial patch size $\ellpatch/=1$ kpc, and in 10 snapshots spanning ${\sim}$ 600 Myr at redshift $z\sim 0$.
        Considered together, (a) and (b) show that, the total ISM pressure is dominated by dispersion/turbulent pressure from the cold+warm gas near the midplane and by thermal pressure in the hot phase at large heights.
        This description of the ISM is independent of the averaging scale \ellpatch/, save for the partitioning of total ``kinetic'' pressure $\pkin/ \equiv \pturb/ + \pflow/$ between the dispersion/turbulence and bulk flow terms.
        }

\section{Vertical Pressure Balance Results}
    \label{s:hydro_balance_results}
    In this section we measure whether (and how) gas achieves vertical pressure balance in different regions of the galaxy. 
    \subsection{Pressure-to-weight ratio maps and scale dependence}
    Figure \ref{f:pressure_ratio_map_scales} maps the total pressure-to-weight ratio, $P_{\rm tot}/W_{\rm tot}$, in a single $z=0$ snapshot of \mtwelvei/ over the face of the disk as a function of \ellpatch/ (in different panels). 
    Each patch is colored according to the
    ${\mathrm{median}_{|z| \leq h}}\left\{\ptot/(z) / W_\mathrm{tot}(z) \right\}$, i.e. 
    the median pressure-to-weight ratio of slabs within $\pm h$ of the midplane.
    Face-on renderings using the same visualization methodology as the top panels of Figure \ref{f:pretty_m12i} are included on the left-hand column to illustrate, e.g., the locations of spiral structure in the disk. 
    Contours of $\Sigma_\mathrm{g} = 10 $ M$_\odot$/pc$^2$ are reproduced in all panels to facilitate comparison between the results for different \ellpatch/~values. 
    The figure shows that, even for averaging scales $\ellpatch/=1$ kpc, $P_{\rm tot}/W_{\rm tot}$ fluctuates substantially from place to place within a given galaxy. 
    Although there are some apparent correlations between gas surface density contours and the pressure ratio (e.g., the inner galaxy, where $\Sigma_\mathrm{g}$ is large, generally has $P_{\rm tot}/W_{\rm tot}<1$), \emph{overall the pattern is complex and spiral features can correspond to either ``over-pressurized'' or ``under-pressurized'' regions}.
    Outside the central region, gas near the midplane typically has a pressure-to-weight ratio $\leq 1$. 
    
    The fluctuations shown in Figure \ref{f:pressure_ratio_map_scales} highlight the extent to which local box simulations, which typically cover disk patches of size $\leq 1$ kpc, cannot capture the full range of conditions occurring throughout a galaxy's ISM.
    Next, we analyze more quantitatively how the pressure-to-weight ratio varies as a function of galactocentric radius and height from the midplane.
    
    \subsection{Pressure-to-weight ratio vs. $R$ and $z$}
    \label{sec:PW_vs_r_z}
    The top row of Figure \ref{f:HB_scatter_vs_radius} shows the height-median pressure ratio within $\pm h$, more precisely defined as 
    $\mathrm{median}_{|z| \leq h} \left\{ \mathrm{mean}_{R,\phi,t} (P_{\rm tot})/ \mathrm{mean}_{R,\phi,t} (W_{\rm tot}) \right\}$,
    for different values of \ellpatch/ and the three simulations analyzed in this paper. 
    This formula indicates that, for fixed $z/h$, we first average over patches within a given annulus ($R$, $\phi$), including patches from snapshots at different times ($t$) in each annulus. 
    Then, for each annulus, we report the median for all slabs with $|z| \leq h$. 
    Overall, the height-median pressure ratio within $\pm h$ is between 0.5-1 for all three simulations at all radii plotted and systematically increases with radius, independent of the averaging scale \ellpatch/.

    The bottom row of Figure \ref{f:HB_scatter_vs_radius} shows a measure of scatter in the pressure ratio vs. radius and \ellpatch/, which we define as ${\mathrm{median}_{|z| \leq h}} \left\{ {\mathrm{IQR}_{r,\phi,t}}(P_{\rm tot} / W_{\rm tot}) \right\}$, 
    where $\mathrm{IQR}$ is the interquartile range (the distance between the $25\th/$ and $75\th/$ percentiles). 
    This may be viewed as the ``height-median of the patch-to-patch scatter within $\pm h$.''
    We find that the scatter increases both as a function of radius and as \ellpatch/ is decreased for all simulations. 
    These trends are, at least qualitatively, consistent with scale-dependent sampling effects \citep[e.g.,][]{Torrey2017,2020MNRAS.493L..87H}. In particular, fewer star-forming complexes are contained within each patch on average with increasing radius and decreasing patch size.
    
    In addition to measuring the height-median within $\pm h$, we also examine vertical profiles of the pressure ratio for three different, large radial annuli ($R=0-3,~3-6,~6-9$ kpc) in Figure \ref{f:HB_vertical_profiles_sims}. 
    In this figure, the vertical profiles show the ratio of means, i.e. $\langle P_{\rm tot} \rangle / \langle W_{\rm tot} \rangle (z) = [\mathrm{mean}_{r,\phi,t} (P_{\rm tot}) / \mathrm{mean}_{r,\phi,t} (W_\mathrm{tot})](z)$, which the derivation in \S \ref{sec:generalizedz_balance} shows should be unity if the system achieves a steady state and the momentum source $\langle \rho S_{z} \rangle$ term averages to zero everywhere. 
    Overall, we find that the $\langle P_{\rm tot} \rangle / \langle W_{\rm tot} \rangle$ ratio is typically within $\approx 20$\% of unity for most heights $|z|$ up to $10h$, in agreement with hydrodynamic pressure gradients on average roughly balancing the weight of the overlying gas. 
    There is only a small dependence on the averaging scale \ellpatch/ in the inner regions where there is only ${\sim}1-10$ patches with \ellpatch/=3 kpc.
    
    A notable systematic departure from $\langle P_{\rm tot} \rangle / \langle W_{\rm tot} \rangle \approx 1$ is found within a disk scale height in most panels: near the midplane, the hydrodynamic pressure is systematically below the weight of the overlying gas, as seen in Figure \ref{f:HB_scatter_vs_radius}.  
    Examining the approximations made in the balance equation derivation (\S \ref{e:general_Euler}), we identify a few possible contributions to these midplane ``pressure dips.'' 
    
    First, $\langle P_{\rm tot} \rangle / \langle W_{\rm tot} \rangle$ will only tend to unity when the momentum source term $\langle \rho S_{z} \rangle \approx 0$. 
    In the limit of infinite resolution, the momentum from SNe and stellar winds in our subgrid model would be injected from point sources with zero net linear momentum by construction \citep[][]{2018MNRAS.477.1578H}. 
    In practice, the momentum is injected over a set of ${\sim} 32$ neighboring resolution elements. 
    In the presence of a vertical gradient in the distribution of momentum sources, the finite injection volume induces a non-zero $\langle \rho S_{z} \rangle$ contribution. 
    We expect this effect to be more important at lower resolution and, as we show in Appendix \ref{a:resolution}, we indeed find systematically lower pressure-to-weight ratios in coarser resolution simulations. 
    A related effect arises in the subgrid model for radiation pressure, which includes a long-range component \citep[][]{Hopkins2019}. 
    The effects of smooth, large-scale radiation supporting the disk against gravity are also not included in $P_{\rm tot}$ because much of the momentum transferred from the radiation field effectively reduces $\vec{g}$ and is not captured by gas motions. 
    Simple estimates indicate that the total momentum injected in the ISM by SNe is ${\sim} 5 \times$ the momentum injected by direct radiation pressure \citep[e.g.,][]{Faucher-Giguere2013}. 
    Thus, if we do not explicitly account for the momentum deposited in the disk by smooth radiation pressure, the ISM can appear ``under-pressurized'' by ${\sim} 20$\% due to the missing radiation pressure. 
    Second, the derivation in \S \ref{e:general_Euler} neglected some correlations between the gas density and velocity fields, e.g. to set $\langle \rho u_{R} u_{z} \rangle = \langle   u_{R}\rangle \langle \rho u_{z} \rangle=0$ (Equation (\ref{e:factor_vels})), but these terms can be non-zero in general. 
    
    Finally, we note that we perform our analysis on cosmological zoom-in simulations, in which galaxies are still evolving during the analysis period and in which the galaxies can be perturbed by various time-dependent external forces (e.g., satellite galaxies).
    Thus, the galaxies are not in an exact statistical steady state and we expect larger departures from unity of the pressure ratio than in more idealized studies based on local boxes (or isolated galaxies), which can achieve equilibrium in a more controlled setting.
    
    Overall, we find that on average vertical hydrodynamic pressure gradients in the simulations analyzed balance the weight of the overlying gas within tens of percent, out to $|z| \geq 10 h$. 
    In particular, the deviations from perfect average equilibrium are much smaller than the scatter in the KS relation (${\sim} 0.5-1$ dex), which equilibrium models are sometimes invoked to explain \citep[e.g.,][]{Faucher-Giguere2013,Hayward2017,Orr2018, Orr2019a}.
    
    \subsection{Pressure contributions from different phases and forms}
    Next, we analyze how the total ISM pressure is decomposed into a sum of partial pressures from different gas phases, as well as contributions from the thermal, velocity dispersion, and bulk flow pressure components defined in \S \ref{sec:generalizedz_balance}.  
    We focus on radii $R \leq 9$ kpc, which are representative of the main parts of the galaxies (i.e. excluding the outermost, lower-density regions, where ISM pressure and star formation may be regulated differently, e.g. owing to the cosmic ultra-violet background) and a fiducial patch size $\ellpatch/=1$ kpc.
    
    Figure \ref{f:pressure_partition} decomposes the total pressure into contributions from each gas phase and different components. 
    From the top panel of Figure \ref{f:pressure_ratio}, we see that the total pressure is provided nearly equally by partial pressures from the warm and cold phases within the galactic disk (i.e. within $\pm h$), mostly by the warm phase up to a height $|z| \leq 2.5 h$, and by the hot phase for heights above that. 

    The bottom panel of Figure \ref{f:pressure_ratio} shows the ratio between the local pressure in each gas phase $\widetilde{P}^\mathrm{phase}_\mathrm{tot}$ and the total ISM pressure \ptot/. 
    Since for each phase the partial and local pressures are related by a factor of the volume filling fraction, $f_{\rm phase}$ (see Equation (\ref{e:local_pressure})), the curves in the top and bottom panels for each phase are related by the $z-$dependent volume filling fraction. 
    
    We see that, for the most part, the phases are in roughly local \textit{total} pressure equilibrium. 
    As we will discuss below, however, the different phases are in general however not in $\emph{thermal}$ pressure equilibrium owing to large non-thermal contributions. 
    Near the midplane, the local pressure in the hot gas is typically larger than \ptot/, which is expected for hot gas generated by SNe. 
    This hot, shocked gas is initially over-pressurized and drives outflows which escape the disk.   
    The large scatter in the local pressures for all gas phases indicates that substantial local fluctuations in pressure balance between different phases are present. 
    This is consistent with a dynamic ISM continuously reshaped by cooling, gravity, and feedback. 
    
    Figure \ref{f:pressure_term_fractions} decomposes the total pressure in each gas phase into the different components \ptherm/, \pturb/, and \pflow/ as a function of height.\footnote{Note that, per the definition in Equation (\ref{e:local_pressure}), the \emph{fractional} partitioning between different forms of pressure is the same for both the partial and local pressures, as the volume filling fraction cancels.}
    Qualitatively, we find that the cold and warm phases have similar contributions from kinetic energy as a function of height, with \pturb/ dominating near the midplane and \pflow/ becoming increasingly dominant with increasing height.
    As expected, the thermal pressure is negligible in the cold phase; it is also everywhere subdominant in the warm phase. 
    The picture is markedly different for the hot phase, where \ptherm/ dominates at all heights, followed by \pflow/, and with \pturb/ contributing only ${\sim} 2\%$ of the total pressure in the hot phase. 
    The fact that the dispersion and bulk pressure components dominate in the cold and warm phases within the disk (i.e., within a few scale heights) indicates that rough equilibrium between \textit{total} local pressures in different phases does not imply \textit{thermal} pressure equilibrium between the phases. 
    In particular, the thermal pressure in the cold and warm phases are much lower than in the hot phase within the disk.\footnote{Note that results regarding whether different phases are in thermal pressure balance with each other can depend on how the phases are defined. For example, \cite{Vijayan2020} define their warm phase as $5050 \leq T \leq 2\times10^{4}$ K and an ``intermediate'' phase as $2\times10^{4} \leq T \leq 10^{5}$ K. \citeauthor{Vijayan2020} find similar local thermal pressures in the warm and intermediate phases, but this is not necessarily inconsistent with our results since both of these phases defined by \citeauthor{Vijayan2020} would be contained within our warm phase with $10^{3}\leq T\leq 10^{5}$ K.}
    
    \figurenv{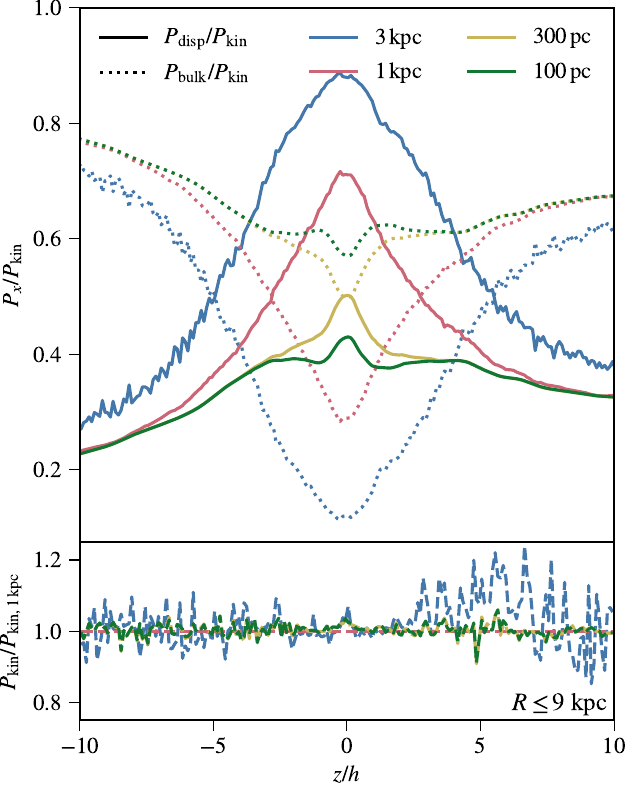}{
        \label{f:pkin_lpatch}
        \emph{Top panel:} Analysis of how the partitioning of the total kinetic pressure $P_{\rm kin}=P_{\rm disp}+P_{\rm bulk}$ (the sum of the dispersion/turbulent and bulk flow components) varies as a function of patch size (or averaging scale) \ellpatch/. 
        The curves show patch-medians for \mtwelvei/ at redshift $z\sim 0$, over a period $\Delta t\approx 600$ Myr, and include all patches with $R \leq 9$ kpc using the fiducial $\ellpatch/= 1$ kpc. 
        Curves of different color show what fractions of the total kinetic pressure are contributed by \pturb/ (solid) and \pflow/ (dotted) for different values of \ellpatch/ as a function of height. 
        The partitioning of kinetic energy between \pturb/ and \pflow/ varies strongly as a function of \ellpatch/ and $|z|$.  
    	Near the midplane, the kinetic energy is concentrated in the velocity dispersion for large \ellpatch/ and in bulk flows for small \ellpatch/. 
    	These limits correspond to whether \ellpatch/ encompasses or is subsumed by individual turbulent eddies, suggesting that the largest turbulent eddies have a size scale between 1 kpc and 300 pc, comparable to the disk scale height. 
        \emph{Bottom panel:} The \pkin/ profiles relative to the values for $\ellpatch/ = 1$ kpc are plotted as dashed lines. 
        These dashed curves show that the total kinetic pressure is constant with varying averaging scale (modulo stochastic fluctuations), which is expected by construction. 
    	}
    	
    The results shown in Figure \ref{f:pressure_partition} are independent of the averaging scale \ellpatch/, save for the partitioning of total ``kinetic'' pressure $\pkin/ \equiv \pturb/ + \pflow/$ between the dispersion/turbulence and bulk flow terms. 
    Figure \ref{f:pkin_lpatch} shows how this partitioning varies with averaging scale and as a function of height from the midplane. 
    As expected, the total kinetic pressure is independent of averaging scale, but the relative importance of the dispersion and bulk flow contributions depends strongly on the averaging scale.
    
    We expect a transition in the partitioning of \pkin/ at $\ellpatch/ \approx \ell_\mathrm{turb}$, where $\ell_{\rm turb}$ is the scale of the largest turbulent eddies. 
    For $\ellpatch/ \gtrsim \ell_\mathrm{turb}$, a single patch contains all scales relevant to the turbulent cascade.  
    On the other hand, for $\ellpatch/ \lesssim \ell_\mathrm{turb}$ the largest turbulent eddies cannot be contained in individual patches. 
    In this limit, the larger turbulent eddies will appear as coherent bulk flows on the scale $\ellpatch/$, and will thus contribute to $\pflow/$ instead of $\pturb/$.
    Near the disk midplane, Figure \ref{f:pkin_lpatch} shows that $P_{\rm kin}$ transitions from being dominated by $\pturb/$ to being dominated by $\pflow/$ as $\ellpatch/$ decreases from 1 kpc to 300 pc. 
    This indicates that the largest turbulent eddies (which contain most of the kinetic energy in a 3D turbulent cascade) have a scale $300\,{\rm pc} \lesssim \ell_\mathrm{turb} \lesssim 1$ kpc. 
    This is consistent with the idea that the largest turbulent eddies have a scale comparable to the scale height of the gas disk (see Figure \ref{f:radial_profiles}).
    As we show in Appendix \ref{a:resolution}, these results are not very sensitive to simulation resolution near the midplane.
    However, we find that at larger heights ($|z|/h \gtrsim 2$), where the gas densities drop substantially and the spatial resolution of our quasi-Lagrangian simulations degrades, the partitioning of the total kinetic pressure between the dispersion and bulk components becomes more sensitive to resolution.
    
    \begin{figure}
        \vspace{-.1225in}
        \centering
        \includegraphics[width=\linewidth]{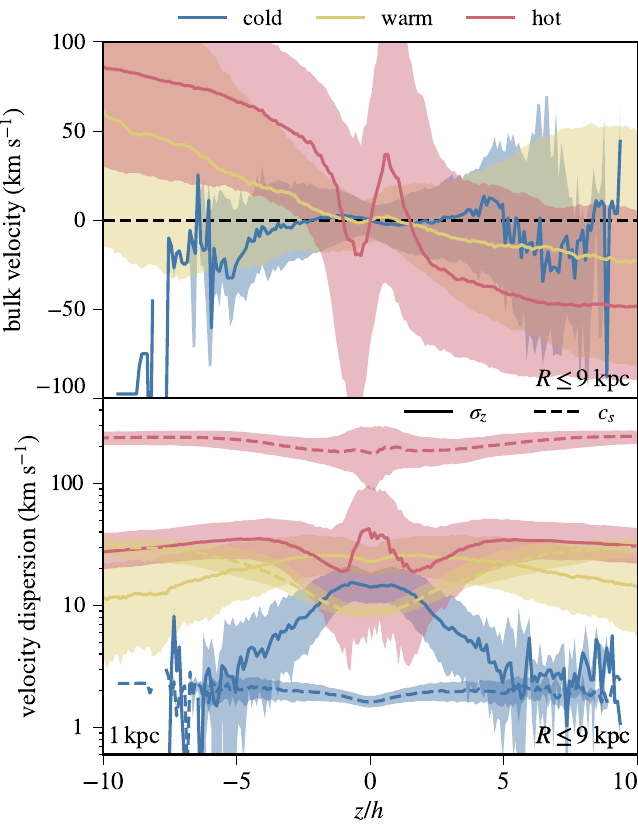}
        \caption{
            \label{f:velocity_profiles} 
            \emph{Top:} vertical bulk velocity ($u_{z}^{\rm s} = \langle u_z \rangle_{\rm M}$) of each of the cold ($T\leq 10^3$ K), warm ($10^3 \leq T \leq 10^5$ K), and hot ($T\geq 10^5$ K) gas phases. 
            The solid curves show patch-medians for \mtwelvei/ in patches with $R\leq 9$ kpc, using a fiducial patch size $\ellpatch/=1$ kpc, and in 10 snapshots spanning ${\sim}$ 600 Myr at redshift $z\sim 0$. 
            Shaded regions are the interquartile range and are shown to quantify the patch-to-patch scatter.
            Positive values correspond to outflowing gas above the     midplane ($z/h > 0$) and inflowing gas below the midplane (and vice versa for $\langle u_z \rangle_{\rm M}<0$).
            The hot gas is outflowing near the midplane and inflowing at larger heights. 
            The same is true for the warm phase, on average, albeit at lower velocity.
            The cold phase, by contrast, has a \textit{median} velocity ${\sim} 0$ at all heights with substantial, ${\sim} \pm 10$ km s$^{-1}$, patch-to-patch scatter.
            \emph{Bottom:} $z-$velocity dispersion and thermal sound speed for each gas phase.
            In both panels, quantities are evaluated for \mtwelvei/ in patches with $R\leq 9$ kpc, using a fiducial patch size $\ellpatch/=1$ kpc, and in 10 snapshots spanning ${\sim}$ 600 Myr at redshift $z\sim 0$.
            By comparing the bulk velocities to the velocity dispersions, we see that the cold gas can be described as being dominated by supersonic turbulence in the disk, the warm gas by supersonic turbulence in the disk and transonic inflows/outflows at larger heights, and the hot gas by subsonic bulk inflows/outflows.}
    \end{figure}

    \doublefigenv{connecting the gas and SFR surface densities through the midplane pressure \label{f:predictors}}{0.625}{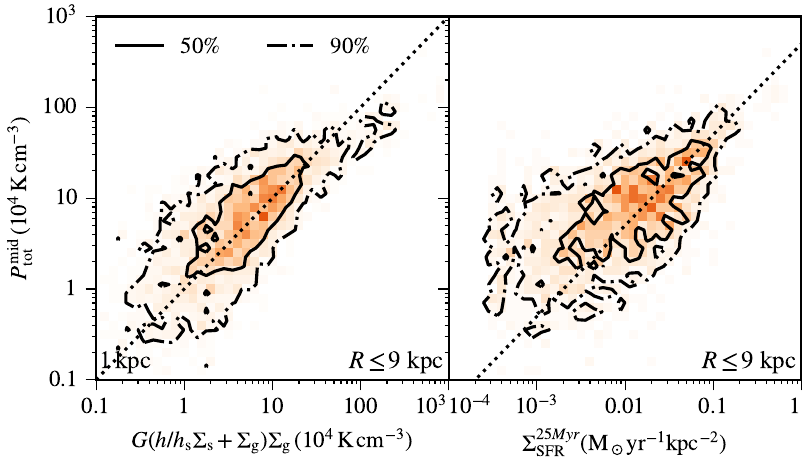}{
        Total midplane pressure (averaged within $|z| \leq h/8$) vs. (left-hand panel) an analytic predictor for the weight of the disk gas ($G(h/h_\mathrm{s})\Sigma_\mathrm{s} + \Sigma_\mathrm{g}) \Sigma_\mathrm{g}$) and (right-hand panel) the star formation rate surface density averaged over 25 Myr. 
        Contours are drawn containing 50\% (solid) and 90\% (dot-dashed) of the patches. 
        A 1-1 line is plotted for reference in the left panel and a line corresponding to $\ptot/^\mathrm{mid} = \sigmasfr/ (1000\mathrm{\, km \, s}^{-1})$ is plotted in the right-hand panel. 
        }
        {components of the weight \label{f:weight_terms}}{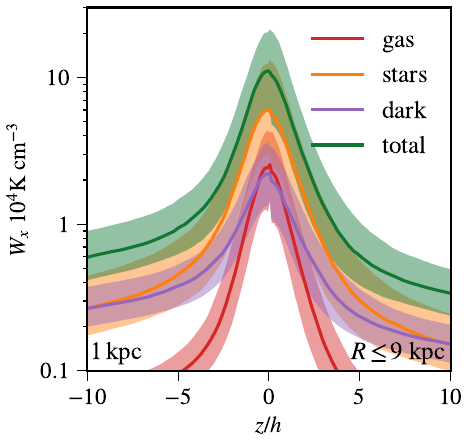}{
        Vertical profiles of the different gas weight components $W_{x}$, defined as in Equation (\ref{e:hb_above}) but in each case including only the contribution to the gravitational acceleration $\vec{g}$ from the gas, stars, or dark matter. 
        The total weight of the disk gas is also shown in green for comparison. 
    	Solid lines are the patch-medians and shaded regions show the inter-quartile range to give an indication of the patch-to-patch scatter.
    	The weight from the stars dominates at all heights, consistent with the low gas mass fractions $f_\mathrm{g} \sim 0.2 \ll 1$ (see Table \ref{t:sim_table}). 
    	In all panels, quantities are evaluated for \mtwelvei/ in patches with $R\leq 9$ kpc, using a fiducial patch size $\ellpatch/=1$ kpc, and in 10 snapshots spanning ${\sim}$ 600 Myr at redshift $z\sim 0$
        }
        
    \subsection{Vertical profiles of bulk velocities and velocity dispersions}
    Lastly, we analyze how the bulk velocity and the velocity dispersion of the gas in the different phases vary as a function of height from the midplane.
    
    The top panel of Figure \ref{f:velocity_profiles} shows how the bulk vertical velocities (corresponding to the slab-averaged $u_{z}^{\rm s}$ defined in \S \ref{sec:generalizedz_balance}) vary as a function of $|z|/h$, measured in 1 kpc patches. 
    We see, for the cold gas, there is little to no vertical bulk motion within a few disk scale heights that is coherent throughout the disk.
    However, the significant fraction of the total pressure contributed by \pflow/ near the midplane in Figure \ref{f:pressure_term_fractions} indicates that while the \textit{patch-median} velocity is close to zero, the $z-$component of the cold gas velocity must fluctuate between positive and negative values between different analysis patches. 
    There is some outflowing cold gas at $|z|/h \gtrsim 3$ but with large scatter in the bulk velocity due to the small amount of cold gas present at these heights. 
    Similarly, there is little bulk motion in the warm gas near the midplane. 
    Interestingly, both the warm gas and the hot gas are on average \emph{inflowing} from $|z| \sim 10h$ to $|z| \sim 2h$, presumably corresponding to cooling of the hot atmospheres surrounding the ${\sim} L^{\star}$ galaxies that we analyze  \citep[e.g.,][]{2019MNRAS.488.2549S, 2020MNRAS.494.3581H}. 
    Some of the inflowing gas can also correspond to the recycling of gas previously ejected by winds \citep[][]{2010MNRAS.406.2325O, 2017MNRAS.470.4698A}. 
    
    Taking into account the full distribution of bulk velocities, all phases contribute to some degree to outflows. 
    The median hot gas bulk velocity shows the most complex behavior.  
    In and around the disk ($|z| \lesssim 2h$), the hot gas is primarily outflowing while at larger heights the bulk velocity changes sign, i.e. the hot gas becomes predominantly inflowing. 
    Overall, the results are consistent with weak ``fountain flows'' in which winds leaving the disk do not reach far into the CGM before turning around or being incorporated into a circumgalactic cooling flow at larger heights. 
    Such weak winds in low-redshift ${\sim} L^{\star}$ galaxies contrast with the much faster and more highly mass-loaded outflows found around the high-redshift progenitors of the same galaxies in FIRE (\citeauthor{Muratov2015} \citeyear{Muratov2015}; \citeauthor{Stern2020} \citeyear{Stern2020}; Chan et al., in preparation).

    The bottom panel of Figure \ref{f:velocity_profiles} shows the $z-$velocity dispersion $\sigma_{z}$ and thermal sound speed $c_{\rm s}$ as a function of height in each of the three different phases. 
    By comparing the bulk velocities to the velocity dispersions in the top vs. bottom panels of the figure, we see that the cold gas can be described as being dominated by supersonic turbulence in the disk, the warm gas by supersonic turbulence in the disk and transonic inflows/outflows at larger heights, and the hot gas by subsonic bulk inflows/outflows.

    The total velocity dispersion profiles (${\sim} \max(c_{s},~\sigma_{z})$) are roughly constant with height within the disk ($|z|/h \lesssim $ 2-3). 

\section{Discussion}
    \label{s:discussion}
    \subsection{The Toomre $Q$ parameter}
    Self-regulation to marginal gravitational stability is a common assumption in models of galactic disks \citep[e.g.][]{Thompson2005,2010ApJ...724..895K,Faucher-Giguere2013,Hayward2017,Orr2018}. 
    The ``classic'' single-component $Q$ parameters for purely stellar or purely gaseous disks \citep[][]{Toomre1964, 1965MNRAS.130..125G} are defined such that local axisymmetric perturbations in infinitely thin, homogeneous disks are gravitationally unstable when $Q \lesssim 1$. 
    Marginal gravitational stability then predicts that disks should have $Q\approx1$. 
    For our adopted definition of $Q_{\rm eff}$, which explicitly includes a stellar component and a gaseous component (see Equation (\ref{e:Qeff})), the criterion for gravitational instability is slightly modified. 
    In an extended stability analysis taking into account finite disk thickness and dissipation, \cite{Elmegreen2011} showed that instability is expected in a two-component stars+gas disk when $Q_{\rm eff} \lesssim 0.75$. 
    
    In \S \ref{s:disk_characteristics} we noted that our disks have an effective Toomre $Q_{\rm eff}$ parameter that is nearly a constant ${\sim} 3$ throughout the main parts of the disk. 
    At face value, since $Q_{\rm eff}$ is well above the nominal \cite{Elmegreen2011} stability threshold, this suggests that the perturbations in the disks should be stable against gravitational collapse. 
    This is however clearly not the case, since the disks are continuously forming stars and in the simulations, star formation is restricted to locally self-gravitating gas. 
    This implies that even more detailed, multi-component analytic disk stability analyses such as that performed by \cite{Elmegreen2011} are not in general sufficient to accurately model gravitational instabilities in realistic disk galaxies. 
    This is not too surprising, since in realistic galaxies the multiphase ISM is continuously stirred and compressed by both gravitational and stellar feedback processes, and dissipation occurs through a complex combination of gas cooling and kinetic dissipation. 
    This leads to highly non-linear perturbations whose development cannot be accurately modeled by a linear analysis of a smooth disk model. 
    
    We note that the relatively large $Q_{\rm eff} \sim 3$ that we find in the simulations goes in the direction that might be expected given the complexities just described. 
    Namely, even though we find $Q_{\rm eff} \sim 3$ when evaluating disk properties averaged in annuli, local gas densities (e.g. in spiral arms) can be much higher than average and thus lead to local gravitational instability and star formation.
    At the same time, large portions of the disk are ``over-supported'' (by e.g. superbubbles and outflows) and do not form stars. 
    In a different analysis of FIRE-2 simulations, \cite{Orr2019} measured $Q \sim 1$ in ${\sim} 1$ kpc patches, but their analysis was focused on regions with non-zero SFR, whereas we include all gas in the present analysis. 
    Although the analysis of \cite{Orr2019} used a different expression for $Q$, we verified that this is a subdominant effect.
    Interestingly, \cite{Leroy2008} analyzed a sample of nearby, observed spiral galaxies and inferred values $Q_{\rm eff}\sim 2-3$ consistent with those found in our simulations. 
  
  \subsection{Vertical pressure balance as an explanation for the KS relation}
\label{s:pism_mid}

    As discussed in the introduction (\S \ref{sec:introduction}), a basic assumption of equilibrium models for the KS relation is that galactic disks are in a state of approximate vertical hydrostatic balance, in the sense that the total ISM pressure near the midplane balances the weight of the overlying disk gas. 
    The results of the previous section show that this requirement is satisfied for the long-lived, low-redshift, ${\sim} L^{\star}$ simulated galactic disks analyzed in this paper. 
    In particular, the results indicate that near the midplane most of the ISM pressure is provided by dispersive motions in the cold and warm phases, which in observations would typically be identified as supersonic or transonic turbulence. 
    The equilibrium is quasi \emph{hydrostatic} near the midplane in the sense that large-scale bulk flows are subdominant.
    
    Vertical hydrostatic balance is however not sufficient on its own to explain the KS relation; for this, there must be a mechanism that couples $\Sigma_{\rm SFR}$ to $\Sigma_{\rm g}$ (or the gas surface density in a specific component, such as molecular gas). 
    In feedback-regulated models, it is postulated that midplane ISM pressure is predominantly provided by momentum injection by stellar feedback. 
    In this case, under simple assumptions for turbulence dissipation, the turbulent pressure $P_{\rm disp} \sim \Sigma_{\rm SFR} (p_{\star}/m_{\star})$, where $p_{\star}/m_{\star}\sim 1000$ km \,s$^{-1}$ is the momentum injected into the ISM by SNe\footnote{There is a factor ${\sim}3$ uncertainty in the correct normalization of $p_{\star}/m_{\star}$ owing not only to uncertainties in the properties of SNe but also where they explode in the ISM and the degree to which the linear momentum from different SNe cancel in the disk.} \citep[e.g.,][]{Ostriker2011, Faucher-Giguere2013, Martizzi2015,Hayward2017}. 
    
    Figure \ref{f:predictors} examines in more detail how, in the simulations analyzed, $\Sigma_{\rm SFR}$ and $\Sigma_{\rm g}$ are consistent with being coupled through the midplane pressure as in feedback-regulated, equilibrium KS models. 
    In the left-hand panel, we show the total midplane pressure $\ptot/^\mathrm{mid}$ in \mtwelvei/ versus the analytic estimate $G \Sigma_\mathrm{disk} \Sigma_\mathrm{g} = G((h/h_s)\Sigma_\mathrm{s} + \Sigma_\mathrm{g}) \Sigma_\mathrm{g}$ for the weight of the overlying gas \citep[such a weight-based expression is often how the midplane pressure is estimated in observations, e.g.][]{Wong2002,Leroy2008,Herrera-Camus2017}.
    The factor $h/h_s$ introduced in the above expression for $\Sigma_\mathrm{disk}$ accounts for the fact that, when the stellar disk is thicker than the gas disk, only the stars with $|z| \lesssim h$ (the gas scale height) contribute to the net gravity acting on the gas component (assuming plane-parallel geometry). 
    Consistent with the smooth scale height radial profiles shown in Figure \ref{f:radial_profiles}, neglecting the $h/h_s$ factor produces only a modest offset and additional scatter in the left-hand panel of Figure \ref{f:predictors}, so it is not critical for the present discussion.
    In the right-hand panel, we show how $\ptot/^\mathrm{mid}$ scales with $\Sigma_{\rm SFR}$ averaged over a time-scale of 25 Myr (this roughly corresponds to the time-scale needed for most of the feedback energy from SNe to be deposited into the ISM).
    
    Consistent with the key results of \S \ref{s:hydro_balance_results} showing that the total midplane pressure is determined by the weight of the overlying gas, the left-hand panel of Figure \ref{f:predictors} shows a clear linear correlation, with approximately the same normalization, for $\ptot/^\mathrm{mid}$ vs. $G \Sigma_\mathrm{disk} \Sigma_\mathrm{g}$. 
    On the right, we see that the midplane pressure also correlates with $\Sigma_{\rm SFR}$, with a normalization $\ptot/^\mathrm{mid} \sim \Sigma_{\rm SFR} (p_{\star}/m_{\star}) $ consistent with $p_{\star}/m_{\star} \sim 1000~{\rm km s^{-1}}$. 
    These scalings are qualitatively similar to those recently observed in \cite{Sun2020} (compare to their Figure 8). 
    They also agree with the basic assumptions of feedback-regulated models and, when combined by eliminating the midplane pressure, imply a KS-type relation between $\Sigma_{\rm SFR}$ and $\Sigma_{\rm g}$.
    KS-type relations are indeed realized in the FIRE simulations \citep[][]{Hopkins2014, Orr2018}, though as discussed at length by \cite{Orr2018}, a quantitative comparison with observations requires a detailed treatment of gas tracers.
    
    The different contributions to the total gravitational acceleration in the disk are plotted in Figure \ref{f:weight_terms}. 
    The figure shows that, for the present simulations, the total gravitational acceleration is dominated by stars, implying that $\ptot/^\mathrm{mid} \propto G \Sigma_{\rm disk} \Sigma_{\rm g} \sim G \Sigma_{\mathrm{s}} \Sigma_{\rm g}$. 
    In more gas-rich galaxies with gas mass fraction $f_{\rm g} \sim 1$, however, the scaling can approach $\ptot/^\mathrm{mid} \propto G \Sigma_{\rm g}^{2}$, which would result in a steeper scaling of $\Sigma_{\rm SFR}$ vs. $\Sigma_{\rm g}$. 
    
    In this discussion of the KS relation, we do not explicitly account for cosmic rays and/or magnetic fields (not included in the simulations we analyze in this paper). 
    \cite{Chan2019_CRs_isolated} ran and analyzed similar but non-cosmological simulations of $\sim L^{\star}$ galaxies (i.e., idealized simulations of isolated galaxies), also including the base FIRE-2 stellar feedback physics. 
    In addition to the base FIRE-2 physics, \cite{Chan2019_CRs_isolated} ran simulations including magnetic fields and different treatments of cosmic ray transport. 
    For cosmic ray transport parameters calibrated to $\gamma-$ray observations of nearby galaxies, \cite{Chan2019_CRs_isolated} found that cosmic ray pressure had weak effects \emph{within} galactic disks, but could potentially help accelerate galactic winds or provide important non-thermal pressure gradients in the circumgalactic medium \citep[see also][]{Ji2020_CR_CGM, Hopkins2020_CR_winds}. 
    In a follow-up study based on fully cosmological FIRE-2 simulations including a cosmic ray transport model also calibrated to available observations (corresponding to an anisotropic cosmic ray diffusion coefficient $\kappa = 3\times10^{29}$ cm$^{2}$ s$^{-1}$ and including streaming), Chan et al. (in preparation) directly quantify pressure support by magnetic fields and cosmic rays as a function of distance from the midplane. 
    Chan et al. (in preparation) show that magnetic fields contribute only $\sim (1-{\rm few})\%$ of the total midplane pressure support \cite[see also][]{Su2018_Bfields}. 
    Cosmic rays, while more important, typically contribute only $\sim 10\%$ of the total pressure support within the disk, though cosmic ray pressure gradients become more important at a few disk scale heights. 
    Thus, we do not expect magnetic fields and cosmic rays to substantially change our conclusions regarding the emergence of the KS relation from vertical pressure balance in the ISM. 
    The effects of magnetic and cosmic ray pressures on gas dynamics in FIRE-2 simulations will be explored in more quantitative detail in follow-up studies (Chan et al., in prep; Trapp et al., in preparation).

    \vspace{-0.15in}
    \subsection{Relating the scatter in the pressure ratio to the observed scatter in the KS Relation}
    Observational work has shown that the scatter in the KS relation increases with decreasing averaging scale \citep[][]{Kennicutt2007,Bigiel2008,Bigiel2010,Leroy2008,Leroy2017}. 
    It is tempting to suggest that deviations from pressure balance in our analysis, which also increase in magnitude as \ellpatch/ is decreased, might drive (part of) the deviations from the mean KS relation. 
    
    In the simple feedback-regulated equilibrium framework, the equations from the previous section imply that
    \begin{equation}
    \Sigma_{\rm SFR} \sim \left( \frac{P}{W} \right) \left( \frac{G}{p_{\star}/m_{\star}} \right) \Sigma_{\rm disk} \Sigma_{\rm g},
    \end{equation}
    where we omit the subscript `tot' in the pressure-to-weight ratio $P/W$ for simplicity. 
   
    Following a standard propagation-of-errors procedure for the scatter,
    \begin{equation}
    \label{eq:sigmaPW_propag}
    \left(\frac{\sigma_{\sigmasfr/}}{\sigmasfr/} \right)^2 =
    \left(\frac{\sigma_{P/W}}{P/W}\right)^2 + 
    \left(\frac{\sigma_{\Sigma_\mathrm{disk}}}{\Sigma_\mathrm{disk}} \right)^2 +
    \left(\frac{\sigma_{\Sigma_\mathrm{g}}}{\Sigma_\mathrm{g}} \right)^2 + {\rm o.t.},
    \end{equation}
    where `o.t.' are other terms that may contribute, e.g.  finite sampling of star-forming regions \citep[e.g.,][]{2014MNRAS.439.3239K}, time delays between star formation events and ISM pressure equilibration \citep[e.g.,][]{Orr2019a}, or variations in $p_{\star}/m_{\star}$. 
    This implies that the observed fractional scatter in $\Sigma_{\rm SFR}$ (at fixed $\Sigma_{\rm g}$) must be larger than the predicted fractional scatter in $P/W$ for the model and simulations to be consistent with observations. 
    
    A crude comparison with available data on the scale-dependent scatter of the KS relation in local spiral galaxies similar to the simulated galaxies \citep[e.g. in][]{Kennicutt2007, Bigiel2010, Leroy2017} suggests that the scatter in $P/W$ predicted by the simulations (see Figure \ref{f:HB_scatter_vs_radius}) is consistent with, but not negligible compared to, the total fractional scatter in observationally inferred $\Sigma_{\rm SFR}$. 
    This indicates that time- and space-dependent scatter in the pressure-to-weight ratio can potentially account for a substantial portion of the scatter in the KS relation. 
    Thus, it would be interesting for future work to model in more detail how fluctuations in $P/W$ translate into scatter in the observed KS relation.

    \subsection{Previous studies of vertical pressure balance}
    Our results paint a dynamic picture of the ISM, one where individual patches can instantaneously be out of vertical pressure balance, but after averaging over significant portions of the disk there is an average pressure-to-weight ratio of $\approx 1$. 
    The time-averaged equilibrium between the total ISM pressure and the weight of the overlying gas is broadly consistent with what has been found in previous studies of vertical hydrostatic balance in more idealized simulations of Milky Way-like conditions based on local boxes or non-cosmological simulations of isolated galaxies  \citep[e.g.,][]{Kim2013,Benincasa2016,Benincasa2019TheGalaxies,Vijayan2020}. 
    Our finding that the total ISM pressure near the midplane is predominantly provided by dispersion/turbulence in the cold and warm phases is also consistent with previous simulations (although the cold phase was not always explicitly modeled), as is the fact that the hot gas generated by SNe tends to be over-pressurized in the disk and drive outflows \citep[e.g.,][]{Vijayan2020}. 
    
    One interesting difference between our results and the local-box simulation of \cite{Vijayan2020}, which goes in a counter-intuitive direction, concerns the variability of the SFR in 1 kpc$^{2}$ patches (the size of the box in \citeauthor{Vijayan2020}). 
    Despite the idealized boundary conditions, the SFR in \citeauthor{Vijayan2020}'s solar neighborhood local box experiences bursts with peak-to-trough amplitude exceeding an order of magnitude. 
    In our cosmological simulations, the SFR similar in 1 kpc$^{2}$ patches varies by only a factor ${\sim}2$ even though there are in principle many additional sources of perturbation (e.g. gas flows from different parts of the disk, spiral structure, or inflows/outflows/substructure in the galactic halo). 
    One possible explanation for this difference in SFR burstiness is that the main form of stellar feedback in \citeauthor{Vijayan2020}'s TIGRESS simulations is SNe \citep[][]{Kim2018}, whereas the FIRE-2 simulations also include photoionization and radiation pressure. 
    \cite{Hopkins2019} showed that including these radiative feedback processes smooths galaxy-scale SFRs in time. In particular, ``early'' radiative feedback limits the star formation efficiency of individual molecular clouds, which in turn limits the formation of large clusters of SNe exploding at the same time.
    
    We can also compare to recent observational studies that have sought to address whether galaxies achieve a state of dynamical equilibrium.  
    Recently, \cite{Sun2020} carried out a detailed observational analysis of this question for 28 nearby star-forming galaxies and focusing on the cold molecular gas. 
    Overall, they find very good agreement between the total pressure in the cold gas (dominated by dispersive/turbulent motions, as in our simulations) and the total weight of the overlying gas. 
    \cite{Sun2020} note that in the observations, self-gravity of the gas is important to verify dynamical equilibrium in the densest cold gas. 
    In our simulation analysis, self-gravity is implicitly neglected for substructures on scales $\ll \ellpatch/$, but this is not a large effect because we exclude gas above a density cut of 50 cm$^{-3}$ from our results (though not as a source of gravity, for which all mass is considered; see \S \ref{s:partial_vs_local_P}). 
    Moreover, as summarized in \S \ref{sec:simulations}, self-gravitating gas rapidly turns into stars in the simulations. 
    
\section{Summary \& Conclusions}

    \label{s:conclusion}
    We use the FIRE-2 galaxy formation simulations to investigate the structure and properties of the ISM 
    in the disks of $z\sim0$ Milky Way-mass galaxies. 
    We analyze in particular the degree to which the ISM pressure can be modeled as being in quasi-hydrostatic balance with the weight of the overlying gas, an assumption which is the basis of many models for the structure of disk galaxies \citep[e.g.,][]{1990ApJ...365..544B} as well as models of the origin of the KS relation \citep[e.g.,][]{Thompson2005, Ostriker2011,Faucher-Giguere2013,Hayward2017,Orr2018}.
    We also analyze how the total ISM pressure is partitioned between different gas phases and different forms of pressure (thermal vs. dispersion/turbulence vs. bulk flows). 
    This allows us to test the assumption of pressure balance between different phases, which is central to classic models of the ISM \citep[e.g.,][]{McKee1977}.
    
    Relative to previous analyses of ISM pressure balance in more idealized local-box simulations or in simulations of isolated galaxies \citep[e.g.,][]{Kim2013, Benincasa2016, Vijayan2020}, our study improves on previous work by including the full cosmological context. 
    This includes inflows and outflows in the circumgalactic medium, as well as potential dynamical disturbances from satellite galaxies. 
    Moreover, the FIRE-2 simulations implement a more complete set of stellar feedback processes, including Type II and Ia SNe, stellar winds, photoelectric heating, photoionization, and radiation pressure. 
    
    On average, we find that the total ISM pressure within the galactic disks approximately balances the weight of the overlying gas, as expected for systems in quasi-hydrostatic equilibrium.  
    In more detail, the simulated galaxies exhibit deviations from exact vertical pressure balance with a scatter that increases with decreasing averaging scale (see Figure \ref{f:HB_scatter_vs_radius}). 
    By breaking the ISM pressure into different gas phases and forms, we find that that the total ISM pressure in the disk is dominated by kinetic pressure due to velocity dispersion in the warm and cold phases. 
    
    Observationally, these dominant forms of pressure may be identified as transonic and supersonic turbulence in the warm ionized medium (WIM), the warm neutral medium (WNM), and the cold neutral medium (CNM). 
    This turbulence likely has important implications for models of the thermal instability and chemistry in the ISM \cite[e.g.][]{2014A&A...567A..16S, 2019ApJ...885..109B}. 
    Above a few disk scale heights, the gas pressure is dominated instead by thermal pressure in the hot phase which drives outflows. 
    In our simulations of low-redshift, ${\sim} L^{\star}$ galaxies these outflows are weak and may be described as ``fountain flows.''
   
    We find that while the different gas phases are in rough \emph{total} pressure equilibrium with each other within the disk, they are not in \emph{thermal} pressure balance. 
    In particular, since velocity dispersion pressure dominates in the warm and cold phases, these would appear to be highly under-pressurized if only their thermal pressure were considered. 
    
    A novel aspect of our analysis is the explicit decomposition of the total kinetic pressure between ``dispersion'' and ``bulk flow'' components (see \S \ref{sec:generalizedz_balance}). 
    While total kinetic pressure is independent of averaging scale, the partitioning between the dispersion and bulk flow components depends on the scale of random motions, which can correspond to turbulent eddies, compared to the patch size. 
    In the disk, we find the bulk flow term becomes dominant over the dispersion term as the averaging scale is decreased from 1 kpc to 300 pc (see Figure \ref{f:pkin_lpatch}). 
    This is consistent with the scale of the largest turbulent eddies being comparable to the ISM scale height.
    
    We also quantify the contributions from different galactic components to the total gravitational acceleration. 
    For the Milky Way-mass galaxies analyzed here, which have modest gas fractions $f_{\rm g}\sim0.2$, the gravitational acceleration due to stars is the most important near the midplane, while the gravity due to the dark matter and the self-gravity of the gas are strongly subdominant. 
    Correspondingly, we find that the simple analytic expression $\ptot/^\mathrm{mid} \sim G( (h/h_s)\Sigma_\mathrm{s}+\Sigma_\mathrm{g})\Sigma_\mathrm{g}$ predicts the total midplane pressure reasonably well. 
    
    Our analysis also shows that the ISM pressure scales linearly with the star formation rate surface density, $\ptot/^\mathrm{mid} \propto \sigmasfr/$. 
    Taken in concert, these results can explain why galaxies obey KS-type relations of the form $\sigmasfr/ \propto \Sigma_\mathrm{g}^{b}$ (for exponents $b\approx 1-2$ which depend on how different mass components contribute to gravity, as well as the gas tracer used) as a result of vertical pressure balance in the ISM of disk galaxies. 
    If this is correct, the scale-dependent fluctuations we find in vertical pressure balance provide insight into one source of scatter in the observed KS relation. 
    
    In the future, it will be important to explore the dynamical balance not only in stable Milky Way-like disk galaxies but also in the ISM of galaxies that exhibit highly bursty star formation, such as at high redshift. 
    It will also be important to investigate whether the simulations predict a well defined KS relation in such regimes, and if so, how it might be realized if vertical pressure balance is not sustained. 
    Additionally, the simulations analyzed in this paper neglected some potentially important physics, such as magnetic fields and cosmic rays. 
    The effects of these additional processes on ISM balance will be explored in a separate study (Chan et al., in preparation). 

\section*{Acknowledgements}
The authors thank the anonymous referee whose helpful comments improved the quality of this manuscript.
AG is grateful to Chang-Goo Kim for very useful discussions regarding the analysis of disk equilibrium in simulations; Samantha Benincasa and Victor Robles for their helpful comments that improved the quality of this paper; and Bridget Haas for her unwavering support and encouragement during the course of the project. 
ABG was supported by a National Science Foundation Graduate Research Fellowship Program under grant DGE-1842165 and was additionally supported by the NSF under grants DGE-0948017 and DGE-145000, and from Blue Waters as a graduate fellow which is itself supported by the NSF (awards OCI-0725070 and ACI-1238993).
CAFG was supported by NSF through grants AST-1412836, AST-1517491, AST-1715216, and CAREER award AST1652522, by NASA through grants NNX15AB22G and 17-ATP17-0067, by STScI through grants HST-GO-14681.011, HST-GO-14268.022-A, and HST-AR-14293.001-A, and by a Cottrell Scholar Award from the Research Corporation for Science Advancement. 
AJR was supported by a COFUND/Durham Junior Research Fellowship under EU grant 609412 and by the Science and Technology Facilities Council [ST/P000541/1]. 
MYG, SW, and JS are supported as a CIERA Fellows by the CIERA Postdoctoral Fellowship Program (Center for Interdisciplinary Exploration and Research in Astrophysics, Northwestern University).
TKC was supported by the Science and Technology Facilities Council astronomy consolidated grant ST/T000244/1.
DK was supported by NSF Grant AST-1715101 and by a Cottrell Scholar Award from the Research Corporation for Science Advancement. 
AW received support from NASA through ATP grant 80NSSC18K1097 and HST grants GO-14734, AR-15057, AR-15809, and GO-15902 from STScI; the Heising-Simons Foundation; and a Hellman Fellowship.
Support for SRL was provided by NASA through Hubble Fellowship grant HST-JF2-51395.001-A awarded by the Space Telescope Science Institute, which is operated by the Association of Universities for Research in Astronomy, Inc., for NASA, under contract NAS5-26555.
The Flatiron Institute is supported by the Simons Foundation.
This research was undertaken, in part, thanks to funding from the Canada Research Chairs program.
Numerical calculations were run on the Quest computing cluster at Northwestern University; the Wheeler computing cluster at Caltech; XSEDE allocations TG-AST120025, TG-AST140023, TG-AST130039, and TG-AST140064; Blue Waters PRAC allocation NSF.1713353; and NASA HEC allocation SMD16-7592, SMD-16-7561, SMD-17-1204, SMD-16-7324, and SMD-17-1375.

\section*{Data Availability}
The data supporting the plots within this article are available on reasonable request to the corresponding author. A public version of the GIZMO code is available at \url{http://www.tapir.caltech.edu/~phopkins/Site/GIZMO.html}. 
Additional data including simulation snapshots, initial conditions, and derived data products are available at \url{http://fire.northwestern.edu/data/}.



\bibliographystyle{mnras}
\bibliography{bibliography,bib2} 




\appendix

\appendix
\section{Dependence of the results on analysis parameters}
\label{a:time_variability}
\begin{figure}
    \centering
    \includegraphics[width=3.05in]{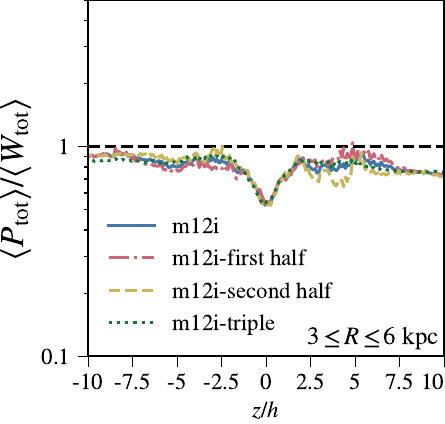}
    \caption{\label{f:time_average_test} Vertical profiles of the annulus-averaged pressure ratio (directly analogous to Figure \ref{f:HB_vertical_profiles_sims}) for two different methods of time averaging in patches with $3 \leq R\leq 6$ kpc, using a fiducial patch size $\ellpatch/=1$ kpc. 
    In the first case, the 10 fiducial snapshots used in the main analysis are split in half, each spanning 300 Myr of the fiducial 600 Myr period (m12i-first half and m12i-second half curves). 
    In the second, $3\times$ as many snapshots are used to span the fiducial 600 Myr period with a spacing of $\approx 20$ Myr between snapshots.
    A reference curve corresponding to the main analysis is provided for comparison (m12i). 
    } 
    
\end{figure}

In Figure \ref{f:time_average_test} we present two tests of how our results depend on the time sampling of snapshots in the simulations. 
The reference result corresponds to the vertical pressure-to-weight profile for the \mtwelvei/ simulation presented in the main text for $\ellpatch/=1$ kpc and $3 \leq R \leq 6$ kpc (see Figure \ref{f:HB_vertical_profiles_sims}). 
For the first test, we split the 10 snapshots used in the main analysis
in two halves, each spanning 300 Myr (the m12i-first half and m12i-second half curves in the figure). 
For the second test, we use $3\times$ as many snapshots over the same total 600 Myr analysis period, with a spacing of $\approx 20$ Myr between snapshots (m12i-triple curve). 
We find that our main results regarding the pressure-to-weight profile are insensitive to the number of snapshots in the time average and the period of time over which the average was performed. 
The results of the same tests for different radial bins are similar.

\begin{figure}
    \centering
    \includegraphics[width=3.05in]{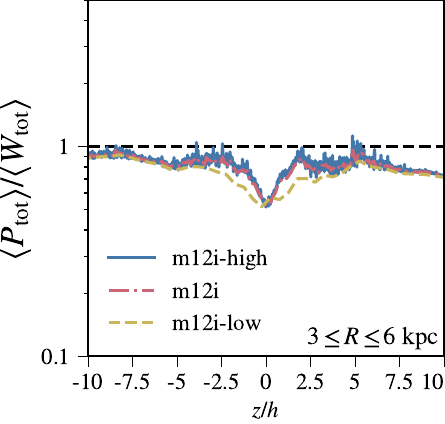}
    \caption{\label{f:vres_test} 
    Similar to Figure \ref{f:time_average_test} but for different analysis slab thicknesses: $0.2\times,1\times,\mathrm{\,and\,} 5\times$ the fiducial slab thickness of $h/10$ (m12i-high, m12i, and m12i-low respectively). 
    The profiles results are well converged for the fiducial slab thickness $\delta z = h/10$ at all radii and heights. 
    } 
    
\end{figure}

In Figure \ref{f:vres_test} we test the dependence of the results on the analysis slab thickness ($\delta z = h/50,h/10,$ and $h/2$). 
We conclude again that the pressure-to-weight profiles are well converged for the fiducial slab thickness $\delta z = h/10$ at all radii and heights considered. 

\section{Deposition of MFM Resolution Elements onto Analysis Grid}

\label{a:sph_smooth}
    Here we describe the method by which we smoothly deposit portions of MFM resolution elements that overlap multiple analysis slabs into those slabs (see Figure \ref{f:disk_schematic}). 
    The procedure is analogous to standard gridding of SPH particle data. 
    
    First, we determine whether a given MFM resolution element indexed by $i$ overlaps multiple slabs by comparing its smoothing length $h_\mathrm{sml}^{i}$ to the grid spacing both horizontally (\ellpatch/) and vertically ($\delta z$). For this purpose, we consider an MFM resolution element to overlap with multiple slabs if and only if
    \begin{equation}
        h_\mathrm{sml}^i > l_{\rm ovlp} \equiv \max\left(\frac{\sqrt{3}}{2}\ellpatch/,~\delta z\right). 
    \end{equation}
    If it does, we assign fractions of the MFM resolution element's mass to overlapping slabs using a cubic spline kernel $w$ such that the mass that resolution element $i$ contributes to slab $s$ is 
    \begin{equation} 
        m_{is} = 
        \begin{cases} 
              m^i \frac{w(r_{is},h_\mathrm{sml}^i)V_\mathrm{slab} }{\sum_{j} w(r_{js},h_\mathrm{sml}^i) V^{j}} 
              \hspace{0.8cm} 
              h_\mathrm{sml}^i > l_{\rm ovlp}
              &\\
      m^i
      \hspace{3.05cm} 
      h_\mathrm{sml}^i \leq l_{\rm ovlp} 
      &\\
        \end{cases},
    \end{equation}
    where $j$ sums over the slabs whose centroid the MFM element's smoothing kernel intersects, $r_{js}$ the distance from the resolution element centroid to the slab center, and $V^{j}=m_{j}/\rho_{j}$ is the effective volume of the MFM element. 
    The sum in the denominator renormalizes the mass deposited to ensure mass conservation. 
    Slabs are then assigned velocity components that are appropriate mass-weighted averages given contributions of overlapping MFM resolution elements. 
    
    The expression for the cubic spline kernel is:
    \begin{equation}
        w(r,h_\mathrm{sml}) = \frac{8}{\pi h_\mathrm{sml}^3}\begin{cases} 
            1-6(r/h_\mathrm{sml})^2 +6(r/h_\mathrm{sml})^3 &\\
            \hspace{2.0cm} 0 \leq r/h_\mathrm{sml} < 0.5&\vspace{0.25cm}  \\
            2(1-r/h_\mathrm{sml})^3 &\\
            \hspace{2.0cm} 0.5 \leq r/h_\mathrm{sml} < 1& \vspace{0.25cm}\\
            0 
            \hspace{1.85cm} 1 < r/h_\mathrm{sml}&
        \end{cases}.
    \end{equation}
    
\section{Dependence of the results on simulation resolution}
\label{a:resolution}

\begin{figure}
    \centering
    \includegraphics[width=3.05in]{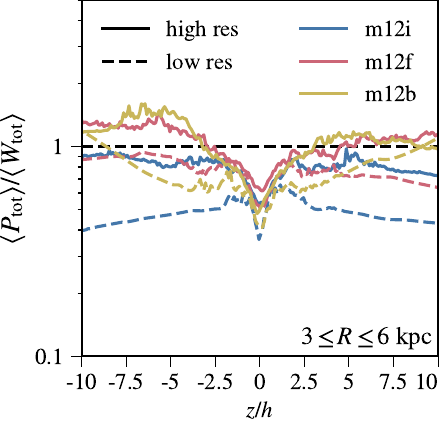}
    \caption{\label{f:resolution_vertical_profile}
    Vertical profiles of the patch-median pressure-to-weight ratio (as in Fig. \ref{f:HB_vertical_profiles_sims} for the $3 \leq R\leq 6$ kpc radial bin) for two different mass resolutions, $m_{\rm b}=7100$ M$_\odot$ (`high res,' as in the main text) and $m_{\rm b}=57000$ M$_\odot$ (`low res'). 
    }
\end{figure}

Here we evaluate how the results presented in the main text depend on the resolution of the hydrodynamic simulations. 
Running simulations at a higher resolution than the state-of-the-art main simulations analyzed in this work ($m_{\rm b}=7100$ M$_\odot$) is computationally prohibitive.
Instead, we compare some of our main results to those from simulations run at $8\times$ coarser mass resolution ($m_{\rm b}=57000$ M$_\odot$).
We find that most of our core results are broadly independent of resolution, but there are some quantitative differences.

Figure \ref{f:resolution_vertical_profile} shows how the patch-median pressure-to-weight ratio profiles depend on simulation resolution for the galaxy models analyzed in this paper, focusing on the radial bin $3\leq R \leq 6$ kpc. 
The results suggest that the pressure-to-weight ratio is systematically lower at coarser resolution, though the differences between the same galaxy model at different resolutions and the differences between between different galaxy models at the same resolution are comparable. 
A systematically lower pressure-to-weight ratio at lower resolution would be consistent with the resolution-dependent effect expected due to the fact that the momentum from supernova and stellar wind feedback is injected over regions of finite size, which scale linearly in radius with the spatial resolution in injection regions (see \S \ref{sec:PW_vs_r_z}). 

\begin{figure}
    \centering
    \includegraphics[width=\columnwidth]{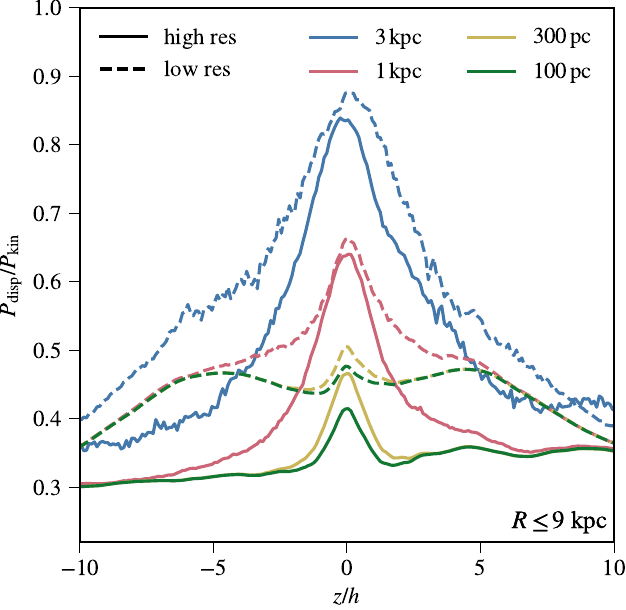}
    \caption{\label{f:resolution_pkin} 
    Ratio of the dispersion pressure to the total kinetic pressure as a function of height from the midplane for \texttt{m12f} vs. \texttt{m12f-low res} ($8\times$ coarser mass resolution), defined as in Figure \ref{f:pkin_lpatch}. 
    }
\end{figure}

Figure \ref{f:resolution_pkin} quantifies how the partitioning of \pkin/ between \pturb/ and \pflow/ depends on simulation resolution, focusing on \texttt{m12f}.\footnote{Unlike for Figure \ref{f:pkin_lpatch} in the main text, we use the \texttt{m12f} galaxy model (rather than \texttt{m12i}) for the resolution test since \texttt{m12i} produces quite different global galaxy properties at the two resolution levels studied \citep[see][]{Hopkins2018}. In this respect, \texttt{m12i} is the exception rather than the norm for FIRE-2 simulations, so the comparison for \texttt{m12f} provides a clearer assessment of hydrodynamic resolution effects.} Since \pkin/=\pturb/+\pflow/ by construction, the partitioning is fully determined by the ratio \pturb//\pkin/. 
The results show that \pturb//\pkin/ is not very sensitive to simulation resolution near the midplane ($\sim5-10$\% differences between resolution levels for a factor of 8 difference in mass resolution). 
However, at larger heights where the gas densities drop substantially and the spatial resolution of our quasi-Lagrangian simulations degrades ($|z|/h \gtrsim 2$), the differences between resolution levels become much larger. The partitioning of the total kinetic pressure between the dispersion and bulk components thus does not appear to be well converged far from the disk midplane.
\bsp	
\label{lastpage}
    
\end{document}